\documentclass[aps, prb, reprint, showpacs, twocolumn, 10pt, groupedaddress]{revtex4-1}

\usepackage{graphicx, amsmath, dsfont, dcolumn, bm, times, color, braket, amssymb}

\begin{document}

\bibliographystyle{apsrev4-1}

\title{Overlapping Andreev states in semiconducting nanowires: competition of 1D and 3D propagation}

\author{Viktoriia Kornich}
\affiliation{Kavli Institute of Nanoscience, Delft University of Technology, 2628 CJ Delft, The Netherlands}
\author{Hristo S. Barakov}
\affiliation{Kavli Institute of Nanoscience, Delft University of Technology, 2628 CJ Delft, The Netherlands}
\author{Yuli V. Nazarov}
\affiliation{Kavli Institute of Nanoscience, Delft University of Technology, 2628 CJ Delft, The Netherlands}

\date{\today}

\begin{abstract}
The recent proposals of devices with overlapping Andreev bound states (ABS) open up the opportunities to control and fine-tune their spectrum, that can be used in various applications. In this Article, we study the ABS in a device consisting of a semiconducting nanowire covered with three superconducting leads. The ABS are formed at two junctions where the wire is not covered. They overlap in the wire where the electron propagation is 1D, and in one of the leads where the propagation is 3D. We identify a number of regimes where these two overlaps either dominate or compete, depending on the junction separation $L$ as compared to the correlation lengths $\xi_{\rm w}$, $\xi_{\rm s}$ in the wire and in the lead, respectively. We utilize a simple model of 1D electron spectrum in the nanowire and take into account the quality of the contact between the nanowire and the superconducting lead.

We present the spectra for different $L$, detailing the transition from a single-ABS in the regime of strong 1D hybridization to two almost independent ABS  hybridized at the degeneracy points, in the regime of weak 1D hybridization. We present the details of merging the upper ABS with the continuous spectrum upon decreasing $L$. We study in detail the effect of quantum interference due to the phase accumulated during the electron passage between the junctions. We develop a perturbation theory for analytical treatment of hybridization.  We address an interesting separate case of fully transparent junctions. We derive and exemplify a perturbation theory suitable for the competition regime demonstrating the interference of 1D and two 3D transmission amplitudes. 
\end{abstract}

\maketitle

\let\oldvec\vec
\renewcommand{\vec}[1]{\ensuremath{\boldsymbol{#1}}}

\section{introduction}
\label{sec:Introduction}
\begin{figure}[tb]
\begin{center}
\includegraphics[width=\linewidth]{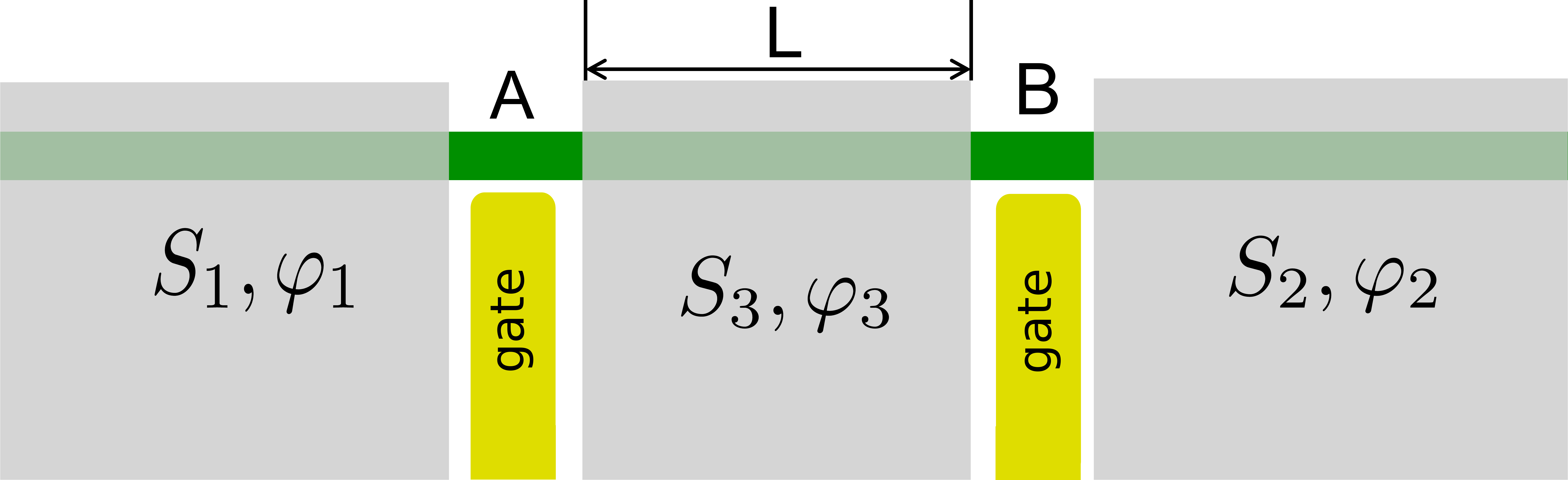}
\caption{The Andreev molecule setup \cite{pillet:nanolett19} consists of a semiconducting nanowire covered by three superconducting leads with the phases $\varphi_1$, $\varphi_2$, and $\varphi_3$. Two junctions $A$ and $B$ are formed in the nanowire. Their transmissions can be tuned by the nearby gates. The ABS at these junctions can be hybridized depending on the separation $L$. }
\label{fig:setup}
\end{center}
\end{figure}

The nanostructures made of semiconducting nanowires in contact with bulk superconducting leads or with superconducting shell are often used in the research aimed to achieve the Majorana-based qubits\cite{lutchyn:prl10, oreg:prl10, mourik:science12, alicea:rpp12}. This boosted the fabrication technology of such nanostructures that has progressed significantly over the last decade\cite{janvier:science15, deacon:prl10, tosi:prx19, su:natcom17, das:natphys12, deng:nanolett12, plissard:nnano13, lee:nnano13, chang:prl13, sherman:nnano16, vanwoerkom:natphys17, goffman:njp17, delange:prl15, larsen:prl15, hays:prl18}. The improved technology makes it possible to realize more sophisticated setups. An interesting setup of an ``Andreev molecule'' has been recently proposed in Ref. \onlinecite{pillet:nanolett19}. In this setup, a nanowire is covered with three superconducting electrodes (Fig. \ref{fig:setup}). The pieces of the nanowire not covered by electrodes form two Josephson junctions. Each junction can host an Andreev bound state (ABS) emerging from the Andreev scattering in the nanowire covered by a superconductor.  If the separation $L$ between the junctions is not too big, these states overlap and hybridize very much like the atomic states do in a molecule, this justifies the term.  Different setups concerning Andreev molecules have been considered  in Refs. \onlinecite{su:natcom17, scheruebl:bjn19, metalidis:prb10}.

We have considered the Andreev molecule setup suggested in Ref. \onlinecite{pillet:nanolett19} in our recent work\cite{kornich:prr19}. We have shown that the energy splitting $\delta E$ at the degeneracy point of two ABS is much smaller than the superconducting gap, $\Delta$. The small parameter involved is an effective resistance of the lead where the ABS overlap, $R$, and $\delta E \simeq \sqrt{R G_{\rm Q}} \Delta$, $G_{\rm Q} \equiv e^2/(\pi\hbar)$ being the conductance quantum. However, this conclusion is based on the assumption of quick electron transfer from the nanowire to the lead. This does not have to be a general case. If the contact between the nanowire and the superconducting lead is not very good\cite{chang:nnano15, kjaergaard:ncom16, guel:nanolett17}, the electrons can stay in a nanowire for a sufficient time to propagate between the junctions without escaping to the lead. In this case, the ABS mainly overlap in the nanowire rather than in the lead, this results in much stronger hybridization\cite{pillet:nanolett19, vanwoerkom:natphys17}. 

\begin{figure}[tb]
\begin{center}
\includegraphics[width=\linewidth]{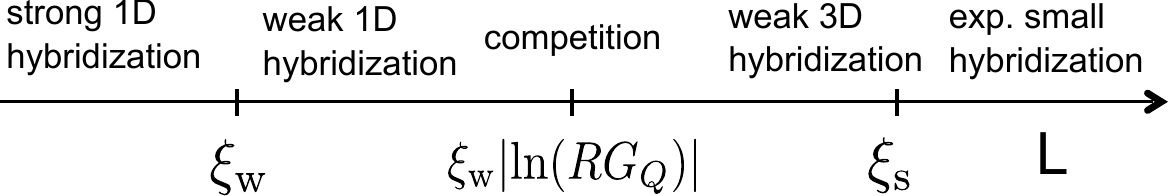}
\caption{The hybridization regimes depending on the junction separation $L$ and the correlation lengths $\xi_{\rm w}, \xi_{\rm s}$ in the nanowire and in the lead, respectively. We distinguish strong 1D hybridization, weak 1D hybridization, competition of 1D and 3D hybridization, weak 3D hybridization. The ABS become independent at $L \gg \xi_{\rm s }$. The 3D case has been considered in Ref. \onlinecite{kornich:prr19}. In this work, we concentrate on the first three regimes.}
\label{fig:regimes}
\end{center}
\end{figure}

In this work, we consider and analyse a number of regimes where 1D or 3D propagation dominate, or the two compete with each other. To characterize the contact between the lead and the nanowire, we use $\tau$, the time a normal electron spends in the nanowire before escaping to the lead (Similar model has been considered in Refs. \onlinecite{sau:prb10, stanescu:prb11}, in their notations, $\tau =\gamma^{-1}$). This gives a correlation length $\xi_{\rm w} = v_{\rm w} \tau$, $v_{\rm w}$ being a typical electron velocity in the wire, that defines a spread of ABS wavefunction in the wire. The condition $L \ll \xi_{\rm w}$ defines the regime of strong 1D hybridization (see Fig. \ref{fig:regimes}). The opposite condition defines the regime of weak 1D hybridization, where the ABS are almost independent except the degeneracy points where they split with $\delta E \simeq \Delta \exp(-L/\xi_{\rm w})$. However, this does not exhaust the regimes. If $\exp(-L/\xi_{\rm w}) \simeq \sqrt{RG_{\rm Q}}$, the overlaps in the wire and in the lead become comparable, and we expect the regime of the competition of 1D and 3D propagation. At further increase of $L/\xi_{\rm w}$, the 3D propagation dominates, this being the case described in Ref. \onlinecite{kornich:prr19}, see Fig. \ref{fig:regimes}. This sequence of regimes implies $\xi_{\rm w} <\xi_{\rm s}$, $\xi_{\rm s}$ being the correlation length in the superconducting lead. The propagation in the lead is naturally diffusive and is characterized by the scattering time $\tau_{\rm s}$, $\xi_{\rm s} \simeq v_{\rm s}\sqrt{\tau_{\rm s}/\Delta}$, $v_{\rm s}$ being the electron velocity in the superconducting material. If the velocities in the superconducting metal and the superconducting wire were the same, the diffusive propagation would have been slower implying $\xi_{\rm w} \ll \xi_{\rm s}$. However, the velocity in the semiconductor is typically two orders of magnitude slower. The condition $\xi_{\rm w} <\xi_{\rm s}$ then implies $\tau\Delta<(v_{\rm s}/v_{\rm w})\sqrt{\tau_{\rm s}\Delta}$. For good contacts between the wire and the superconductor, $\tau \simeq 0.2 \Delta$ \cite{Lutchyn} and the condition holds even for rather dirty superconductors $\tau_{\rm s} \Delta \ll 10^{-4}$.

We investigate the resulting ABS spectrum in all these regimes. Starting from a simple model of 1D semiconducting spectrum augmented with self-energy describing superconducting proximity effect, we derive scattering matrix formalism that permits to compute and understand the ABS energies in 1D regimes. We extend this formalism to include 3D propagation amplitudes to describe the competition regime. We present the spectra for different $L$, illustrating the transition from a strong 1D hybridization regime for $L/\xi_{\rm w}\ll 1$ to the regime with two energy levels with a sizeable splitting at $L/\xi_{\rm w}\sim 1$, and further to almost independent ABS hybridized at the degeneracy points, for $L/\xi_{\rm w}\gg 1$. We present the details on how the upper energy level disappears merging with the continuous quasiparticle spectrum upon decreasing $L$. We study the effect of quantum interference on the spectrum in various regimes, that is, the oscillatory dependence on the phase accumulated during the electron passage between the junctions. We demonstrate that the energies can be significantly affected by the interference for $L/\xi_{\rm w}\ll 1$ in the whole range of the phases, while for larger $L/\xi_{\rm w}$ the interference is pronounced only in the vicinity of the degeneracy points. We provide analytical formulas for this case.  We separately address an interesting case of ballistic junctions and discuss its peculiarities with respect to other results. We derive and analyse analytical formulas for the competition regime demonstrating the interference of 1D and two 3D transmission amplitudes. We show that the variances of 3D amplitudes are the same and scale as $\sim G_{\rm Q}R$. As the 1D transmission amplitudes scale as $e^{-L/\xi_{\rm w}}$, the competition regime occurs, when these two scales are of the same order. We derive an analytical formula for the energy splitting due to 3D propagation and compare it to the results of Ref. \onlinecite{kornich:prr19}.  

The paper is organized as follows. In Sec. \ref{sec:DeviceModel} we present the details of the setup and the model in use. We consider the wave functions and the spectrum edge for the infinite uniform nanowire and discuss the dependence on the parameter $\tau\Delta$ in Sec. \ref{sec:WaveFunctions}. The scattering matrix approach is derived and outlined in Sec. \ref{sec:ScatteringMatrix}. We summarize and discuss the main results in  Sec. \ref{sec:MainResults}. In Sec. \ref{sec:ShortLead} we consider the strong 1D hybridization. We develop a perturbation theory suitable in the opposite limit, in Sec. \ref{sec:LongLead}. The detailed discussion of the interference effect is presented in Sec. \ref{sec:Interference}. The transfer between single-band and two-band regimes is detailed in Sec. \ref{sec:Transition}. The Sec. \ref{sec:Transparent} focuses on the case of the fully transparent junctions. The competition regime is considered in Sec. \ref{sec:Interplay}. We conclude in Sec. \ref{sec:Conclusions}. 
 
\section{The setup and model}
\label{sec:DeviceModel}
Let us detail the Andreev molecule setup (Fig. \ref{fig:setup}). Electrically, this is a three-terminal circuit with two junctions. We assume same superconducting material for all electrodes, so that the superconducting gap is the same for all of them. The spectrum of the bound states will depend on three superconducting phases of the electrodes, $\varphi_1$, $\varphi_2$, and $\varphi_3$. In fact, by virtue of gauge invariance, it depends only on two phase differences $\tilde{\varphi}_1 = \varphi_1 - \varphi_3$, $\tilde{\varphi}_2 = \varphi_1 - \varphi_3$. If the junctions can be regarded as independent, two independent ABS with energies $E_{1,2}(\tilde{\varphi}_{1,2})$ are formed. If the ABS are hybridized, each energy depends on both phase differences. We assume that the wire is sufficiently long in comparison with the electron wavelength, $ k_{\rm F}L \gg 1$.

We describe the electron spectrum in the nanowire with a minimal model. We assume that the nanowire has a single propagation mode, disregard the spin splitting and concentrate on the states close to the Fermi surface. Since the energies of the ABS are of the order of the proximity-induced gap $\tilde{\Delta}$, this implies sufficiently big Fermi energy $E_F\gg \tilde{\Delta}$.  The Hamiltonian with the linearized spectrum is naturally written as a matrix in the basis of right- and left moving electrons, whose field operators are envelope functions of $\exp(\pm k_{\rm F} x)$, $\Psi_\sigma(x) = \exp(ik_{\rm F}x)\Psi_{R,\sigma}(x) + \exp(-ik_{\rm F}x)\Psi_{L,\sigma}(x)$, $x$ being an effective coordinate along the nanowire, $\sigma$ being spin index. It reads:
\begin{eqnarray}
\label{eq:Hnw}
H_{\rm nw}&=&\int dx' dx \sum_{\alpha,\beta={R,L};\sigma} \Psi^\dagger_{\alpha,\sigma}(x') H^{\rm nw}_{\alpha \beta} (x,x') \Psi_{\beta,\sigma}(x),\ \ \ \  \\ \hat H^{\rm nw} &=& -iv_{\rm w} \frac{\partial}{\partial x}\tau_z+\hat{V}_A(x)+\hat{V}_B(x)
\end{eqnarray}
 Here, $v_{\rm w}$ is the Fermi velocity,  $\tau_z$ is a diagonal matrix with $\tau_z^{RR}= -\tau_z^{LL}=1$. We assume that the wire is ballistic under the electrodes while the electrons are scattered in the junction regions, $\hat{V}_A(x)$ and $\hat{V}_B(x)$ are the matrix potentials responsible for this scattering. In principle, there is no much work to generalize $H_{\rm nw}$ and to include parabolic dispersion, spin-orbit splitting and spin magnetic field\cite{tosi:prx19, lutchyn:prl10, oreg:prl10}. However, in this Article, we would like to focus on the phenomenon of hybridization that does not necessarily involve spin, so we keep it simple. The Fermi energy, $v_{\rm w}$ and $k_{\rm F}$ in the nanowire can be changed by the applying voltage to an underlying gate\cite{mourik:science12}. Importantly, even small changes of this gate voltage can cause significant change of the phase $k_{\rm F}L$ accumulated by an electron moving between the junctions.

The Hamiltonian describing the $j$th superconducting lead, where $j=\{1,2,3\}$, is convenient to write not specifying the orbital electron states present in a disordered superconductor. We label these states with $q$, and assume a homogeneous superconducting order parameter $\Delta e^{i\varphi_j}$. In terms of the corresponding creation/annihilation operators $d_{q,\sigma}^\dagger$ and $d_{q,\sigma}$ the Hamiltonian reads as follows:
\begin{equation}
\label{eq:BdGHamiltonian}
H_j=\sum_q \xi_n d_{q,\sigma}^\dagger d_{q,\sigma} +\Delta e^{-i\varphi_j} d_{q,\uparrow}d_{q,\downarrow}+\Delta e^{i\varphi_j}d_{q,\uparrow}^\dagger d_{q,\downarrow}^\dagger.
\end{equation}
$\xi_n$ being the energies of the orbital states counted from the Fermi energy. 

The contact between the nanowire and a lead is of tunneling nature and is described with a  tunneling Hamiltonian 
\begin{equation}
\label{eq:TunnelingHamiltonian}
H_T=\sum_{k,q}t_{k,q}a_{k,\sigma}^\dagger d_{q,\sigma} +t_{k,q}^*d_{q,\sigma}^\dagger a_{k,\sigma},
\end{equation}
 $k$ labeling the normal-electron states in the nanowire, $a_k^\dagger$ and $a_k$ being the creation/annihilation operators in these states. The tunnel coupling $t_{k,q}$ depends on the electron states in both the nanowire and the leads. In the absence of superconductivity, the escape rate from the state $k$ to the lead, $1/\tau_k$ is given by the Fermi Golden Rule 
\begin{equation}
\frac{1}{\tau_k}=\frac{2\pi}{\hbar}\sum_q|t_{k,q}|^2\delta(E_k-\xi_q). 
\end{equation}
It is convenient and realistic to assume that this escape rate does not depend on the state, so the quality of the contact between the nanowire and the leads is characterized by a single escape time $\tau$.

Under these circumstances, the tunneling into a lead can be conveniently incorporated into a local self-energy\cite{sau:prb10, stanescu:prb11} $\Sigma_j$, which is a matrix in the basis of right- and left-moving electrons and holes $(\Psi^{e,R},\Psi^{h,L},\Psi^{e,L},\Psi^{h,R})$
\begin{eqnarray}
\label{eq:SelfEnergy}
\Sigma_j&=&\frac{1}{\tau\sqrt{\Delta^2-E^2}}\begin{pmatrix} -E& \Delta e^{i\varphi_j} & 0 & 0 \\ \Delta e^{-i\varphi_j} & -E & 0 & 0 \\ 0 & 0 & -E & \Delta e^{i\varphi_j}\\ 0 & 0 & \Delta e^{-i\varphi_j} & -E\end{pmatrix},\ \ \ \ \
\end{eqnarray}
so the resulting equation for the Green's function in the nanowire reads:
\begin{eqnarray}
\label{eq:Green}
\left( E - {\cal H} \right) G(x,x') = - \delta(x-x')\\
{\cal H} =  -i v_{\rm w} \eta \frac{\partial}{\partial x}  + W_A(x) + W_B(x) + \Sigma(x),
\end{eqnarray}
with 
\begin{eqnarray}
\eta &=& \begin{pmatrix} 1& 0 & 0 & 0 \\ 0 & -1 & 0 & 0 \\ 0 & 0 &-1 & 0\\ 0 & 0 & 0 & 1\end{pmatrix}, \\
W_A &=& \begin{pmatrix} V^{RR}_{A}& 0 & V^{RL}_{A} & 0 \\ 0 & -V^{LL}_A & 0 & -V^{RL}_A \\ V^{LR}_A & 0 &V^{LL}_A & 0\\ 0 & -V^{LR}_A & 0 & -V^{RR}_{A}\end{pmatrix},
\end{eqnarray}
$W_B$ having the same structure.
\section{Uniform nanowire}
\label{sec:WaveFunctions}

In this Section, we will consider the spectrum and the wavefunctions in an infinite and uniform semiconducting nanowire with the proximity-induced  gap $\tilde{\Delta} <\Delta$. There are no states at energies below $\tilde{\Delta}$ in a uniform nanowire, there are modes confined in the nanowire at $\tilde{\Delta}<E<\Delta$, and there are extended states in the wire and leads at $E>\Delta$. For a uniform wire, we can regard ${\rm det}(E-{\cal H})$ as an equation for the wave vector for a given energy. Correspondingly, the wave vector is imarinary at $0<E<\tilde{\Delta}$, is real in the interval $\tilde{\Delta} <E<\Delta$, and complex otherwise.

\begin{figure}[tb]
\begin{center}
\includegraphics[width=\linewidth]{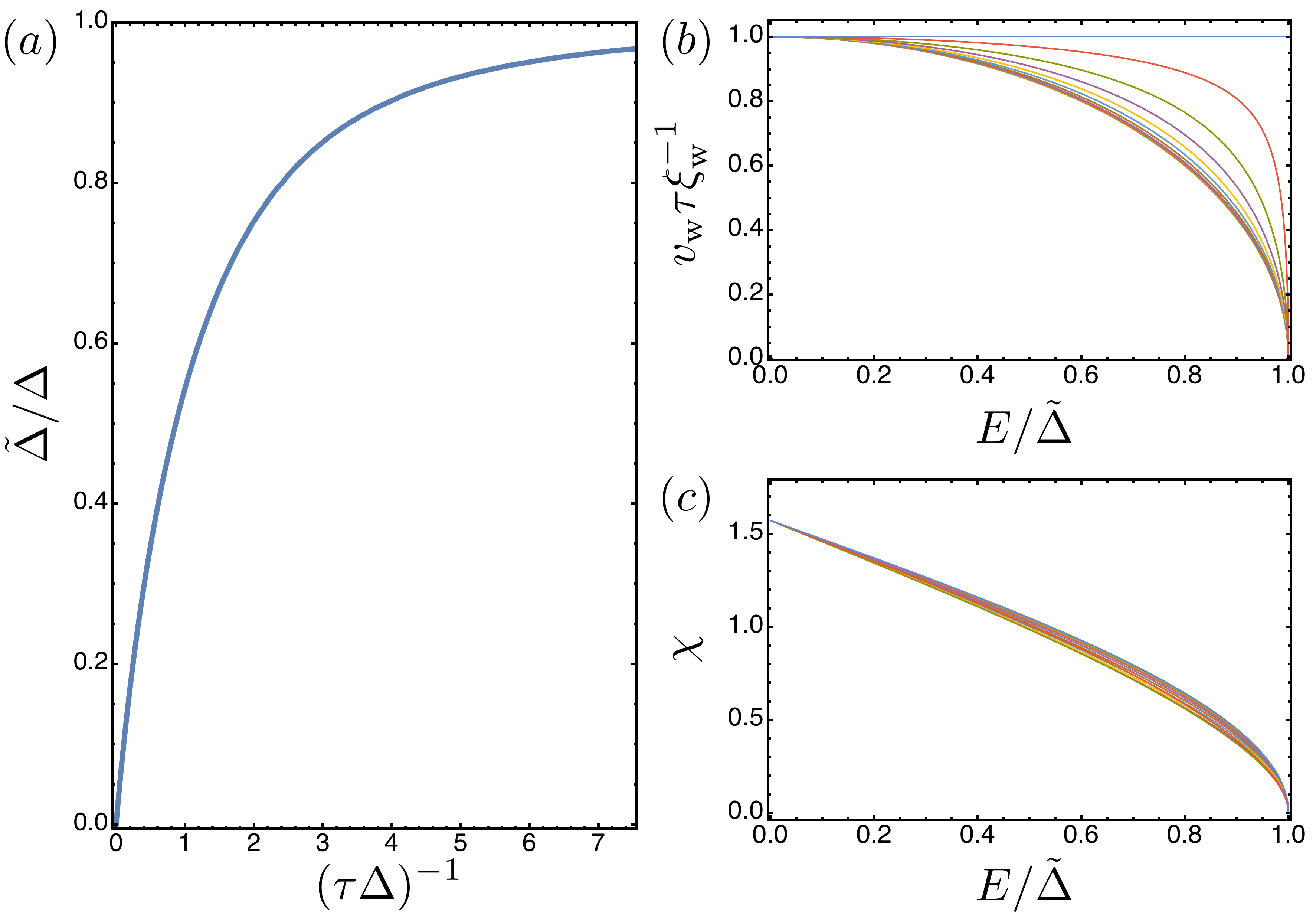}
\caption{(a) The relative proximity gap  $\tilde{\Delta}/\Delta$ versus the parameter $(\tau\Delta)^{-1}$ characterizing the quality of the tunnel contact between the nanowire and the superconducting lead. For a good contact, $\tau\rightarrow 0$, $\tilde{\Delta} \rightarrow \Delta$. (b) The inverse correlation length $\xi_{\rm w}(E)$ versus energy for different values of $\tau\Delta$. (c) The Andreev reflection phase $\chi$ versus energy. In both plots, the values of the parameter for different curves correspond to $\tilde{\Delta}/\Delta=\{0,0.1,0.2,0.3,...,0.9,0.98,1\}$.}
\label{fig:chiE}
\end{center}
\end{figure}

Since we will later concentrate on ABS, we concentrate at $E<\tilde{\Delta}$. The imaginary part of the wave vector gives an energy-dependent inverse localization length $\xi^{-1}_{\rm w}$:
\begin{eqnarray}
\label{eq:kappa}
v_{\rm w} \tau \xi^{-1}_{\rm w}=\sqrt{1-E^2\tau^2-\frac{2E^2\tau}{\sqrt{\Delta^2-E^2}}}.
\end{eqnarray} 
The condition $\xi^{-1}_{\rm w}(E)=0$ eventually defines the gap $\tilde{\Delta}$. It is given by an implicit relation
\begin{equation}
\tau\Delta =\frac{\Delta}{\tilde{\Delta}}\sqrt{\frac{\Delta-\tilde{\Delta}}{\Delta+\tilde{\Delta}}}.
\end{equation}
and is plotted in Fig. \ref{fig:chiE} (a) as a function of $(\tau \Delta)^{-1}$. Short $\tau$ implies a good contact, so $\tilde{\Delta} \approx \Delta$ at $\tau\Delta \ll 1$. In the opposite limit, $\tilde{\Delta} \approx 1/\tau \ll \Delta$. 
In Fig. \ref{fig:chiE} (b) we plot the inverse correlation length versus energy normalized by the proximity gap $\tilde{\Delta}$, for various $\tau\Delta$. We see that for any value of this parameter the correlation length is close to the escape length $v_{\rm w}\tau$. For a bad contact, 
$v_{\rm w}\tau \xi^{-1}_{\rm w} = \sqrt{1-(E/\tilde{\Delta})^2}$, for a good contact $\xi_{\rm w} = v_{\rm w}\tau$ for all energies except the vicinity of the gap edge.

There are four eigenfunctions at each energy, corresponding to right- or left-moving electrons and the exponent decreasing either to the left or to the right, 
\begin{eqnarray}
\label{eq:wavefunctions}
\begin{pmatrix}\Psi^{e,R}\\ \Psi^{h,L}\end{pmatrix}&=&\begin{pmatrix}1 \\ e^{i (\mp\chi-\varphi)}\end{pmatrix}e^{\mp x/\xi_{\rm w}},\ \ \ \ \ \ \\
\begin{pmatrix}
\Psi^{e,L}\\ \Psi^{h,R}
\end{pmatrix}&=&\begin{pmatrix}1 \\ e^{i(\pm\chi-\varphi)}\end{pmatrix} e^{\mp x/\xi_{\rm w}}.\ \ \ \ \ \ \
\end{eqnarray}

Here, we introduce an important phase $\chi$ associated with the phase of Andreev reflection from a corresponding piece of the nanowire,
\begin{eqnarray} 
\label{eq:chi}
\chi=\arcsin\sqrt{1-\left[\frac{E(1+\tau\sqrt{\Delta^2-E^2})}{\Delta}\right]^2}
\end{eqnarray}
in the interval $0<E<\tilde{\Delta}$. As we will see, the ABS energies are determined from the energy dependence of $\chi$. At any value of $\tau\Delta$, $\chi(0)=\pi/2$, $\chi(\tilde{\Delta})=0$. It is interesting to note that $\chi(E/\tilde{\Delta})$ exhibits very little dependence on $\tau\Delta$. This is seen in Fig. \ref{fig:chiE} (c) where all the curves corresponding to different $\tau\Delta$ collapse into one. This is why the ABS spectrum is hardly sensitive to $\tau\Delta$, and we do not have to explore  its dependence on this parameter.

\section{Scattering Approach}
\label{sec:ScatteringMatrix}
To avoid describing the details of the junctions and the corresponding potentials in their vicinity, we implement the scattering approach for the problem under consideration. The scattering approach to the setup was first implemented in Refs. \onlinecite{pillet:nanolett19, pillet:arxive} at lesser detail level. Their results are qualitatively the same. A scattering matrix, by definition, is a matrix that relates the outgoing wave amplitudes to incoming ones. In the setup under consideration, there are two junctions, $A$ and $B$ (see Fig. \ref{fig:setup}). We assume that the junction region is shorter than $\xi_{\rm w}$, this assumption permits to neglect possible Andreev scattering in the junctions as well as the energy dependence of the scattering amplitudes at the energy scale $\simeq \tilde{\Delta}$. If we regard the junction $A$ as a scattering region, the incoming electron wave amplitudes are $\{\Phi_1^{e,R}, \Phi_3^{e,L}\}$ and the outgoing ones are $\{\Phi_1^{e,L}, \Phi_3^{e,R}\}$, where $1,3$ refer to the leads bounding the junction $A$, and the amplitudes correspond to the wave functions on the side of a lead. The electron scattering matrix for the junction $A$ in this basis is
\begin{eqnarray}
S_A^{e}=\begin{pmatrix}r_A e^{- i\theta_1^A} & t_A e^{- i\frac{\theta_1^A+\theta_3^A}{2}}\\ t_A e^{- i\frac{\theta_1^A+\theta_3^A}{2}} & - r_A e^{- i\theta_3^A}\end{pmatrix}.
\end{eqnarray}
Here, real $r_{A}$ and $t_{A}$, $r^2_A + t_A^2 =1$,  denote reflection and transmission amplitudes, $\theta_{1,3}^{A}$ are the corresponding reflection phases. The electron scattering matrix for junction $B$, $S_B^e$, is defined in a similar basis: the incoming amplitudes are $\{\Phi_4^{e,R}, \Phi_2^{e,L}\}$ and outgoing ones are $\{\Phi_4^{e,L}, \Phi_2^{e,R}\}$, where $4$ referes to the wave functions in the lead $3$ close to the junction $B$. The matrix reads:
\begin{eqnarray}
S_B^{e}=\begin{pmatrix}r_B e^{- i\theta_3^B} & t_B e^{- i\frac{\theta_3^B+\theta_2^B}{2}}\\ t_B e^{- i\frac{\theta_3^B+\theta_2^B}{2}} & - r_B e^{- i\theta_2^B}\end{pmatrix}.
\end{eqnarray}
The  scattering matrix for holes is obtained from the electron one via complex conjugation. Thus, the total scattering matrix describing the scattering from the junctions, $S_{\rm NS}$, relates the incoming amplitudes $\Phi_+=\{\Phi_1^{e,R}, \Phi_3^{e,L}, \Phi_1^{h,R}, \Phi_3^{h,L}, \Phi_4^{e,R}, \Phi_2^{e,L}, \Phi_4^{h,R}, \Phi_2^{h,L}\}$ to the outgoing ones $\Phi_-=\{\Phi_1^{e,L}, \Phi_3^{e,R}, \Phi_1^{h,L}, \Phi_3^{h,R},\Phi_4^{e,L}, \Phi_2^{e,R},\Phi_4^{h,L},\Phi_2^{h,R}\}$, and has a block-diagonal form
\begin{eqnarray}
S_{\rm NS}&=&\begin{pmatrix}S_A^e & 0 & 0 & 0\\
0 & S_A^h & 0 & 0\\
0 & 0 & S_B^e & 0\\
0 & 0 & 0& S_B^h
\end{pmatrix}.
\end{eqnarray} 
Andreev scattering occures in the wire regions covered by superconducting leads. The outgoing wave amplitudes for $S_{\rm NS}$  are  incoming wave amplitudes for Andreev scattering matrix $S_{\rm AS}$ and vice versa. This gives $\Phi_-=S_{\rm AS}\Phi_+$, and the matrix $S_{\rm AS}$ is derived from the matching of the wavefunctions (\ref{eq:wavefunctions}). It reads:
\begin{eqnarray}
S_{\rm AS}&=&\begin{pmatrix}
0 & 0 & r_1^{eh} & 0 & 0 & 0 & 0 & 0\\
0 & 0 & 0 & r_3^{eh} & t_{R}^e & 0 & 0 & 0\\
r_1^{he} & 0 & 0 & 0 & 0 & 0 & 0 & 0\\
0 & r_3^{he} & 0 & 0 & 0 & 0 & t_{R}^h & 0\\
0 & t_{L}^e & 0 & 0 & 0 & 0 & r_4^{eh} & 0\\
0 & 0 & 0 & 0 & 0 & 0 & 0 & r_2^{eh}\\
0 & 0 & 0 & t_{L}^h & r_4^{he} & 0 & 0 & 0\\
0 & 0 & 0 & 0 & 0 & r_2^{he} & 0 & 0\\
\end{pmatrix},
\end{eqnarray}
with 
\begin{eqnarray}
\label{eq:scattering_amplitudes}
&&r_{1,2}^{eh,he}=e^{i(\pm\varphi_{1,2}+\chi)},\\
&&r_3^{eh,he}=r_4^{eh,he}=e^{i(\pm\varphi_3+\chi)}r_3,\\
&&r_3=\frac{1-e^{-2L/\xi_{\rm w}}}{1-e^{2i\chi}e^{-2L/\xi_{\rm w}}},\\
\label{eq:transmission}
&&t^{e,h}_R=t^{e,h}_L=e^{\pm ik_{\rm F}L}t,\\
\label{eq:transmissiont}
&&t=\frac{(1-e^{2i\chi})e^{-L/\xi_{\rm w}}}{1-e^{2i\chi}e^{-2L/\xi_{\rm w}}},\\
&&|t|^2+|r_3|^2=1.
\end{eqnarray}
The notations $eh$ and $he$ imply the electron conversion into a hole and vice versa. The transmission amplitudes $t_{R,L}^{e,h}$ do not involve a conversion and correspond to electron or hole propagation through  the part of the nanowire under the third lead. The  phases $\pm k_{\rm F}L$ acquired in the course of propagation are manifested in the quantum interference effect, as we will show later. For a small separation between the junctions, $L/\xi_{\rm w}\ll 1$, $r_3  \to 0$ and $|t| \to 1$. This implies that the electrons or holes do not exhibit Andreev reflection directly passing to another junction. In the opposite limit, $L/\xi_{\rm w}\gg 1$, $|r_3|=1$, and $|t|=0$. The scattering matrix is separated into blocks indicating the separation of ABS formed at the two junctions are completely separated from each other.

Since $\Phi_-=S_{\rm AS}\Phi_+$ and $\Phi_+=S_{\rm NS}\Phi_-$ an ABS is formed provided $S_{\rm NS} S_{\rm AS}$ has a unit eigenvalue. This gives an equation that is satisfied at an energy corresponding to an ABS energy,
\begin{equation}
\label{eq:det}
\det{(1-S_{\rm NS}S_{\rm AS})}=0. 
\end{equation}
In this work, we solve this equation numerically and analytically for various cases.

\section{overview of the ABS spectrum}
\label{sec:MainResults}

In this Section, we discuss the propagation processes in the setup, relate those to the features of the spectrum, and give an overview of the concrete results. To start with, we shall note that the hybridization of ABS states formed at two junctions requires either electron or hole propagation between the junctions. This is evident from the scattering approach where the scattering matrix is separated into the blocks for each junction unless there are non-zero transmission amplitudes $t^{e,h}_{R,L}$. This propagation may naturally take place in 1D wire, or involve an escape to the 3D lead with a subsequent return to the wire.

In the strong 1D hybridization regime $L \ll \xi_{\rm w}$ the propagation between the junctions is unobstructed by anything, even by Andreev reflection, since the propagation time is too short for a particle to feel the induced gap in the nanowire. As the result, the third electrode has no effect on the ABS, and we have a compound junction between $A$ and $B$ that supports a single ABS. We show this explicitly and analytically in Sec. \ref{sec:ShortLead}. In the opposite limit $L \gg \xi_{\rm w}$ of the weak 1D hybridization the direct propagation is strongly reduced by Andreev reflection in the wire: an electon/hole is turned back as a hole/electron. There are two independent ABS and hybridization is only important in the vicinity of degeneracy points where two energies cross. We develop a perturbation theory valid for a small direct transmission amplitude (Sec. \ref{sec:LongLead}) that provides an analytical expression for this splitting for general scattering matrices. 

\begin{figure}[tb]
\begin{center}
\includegraphics[width=\linewidth]{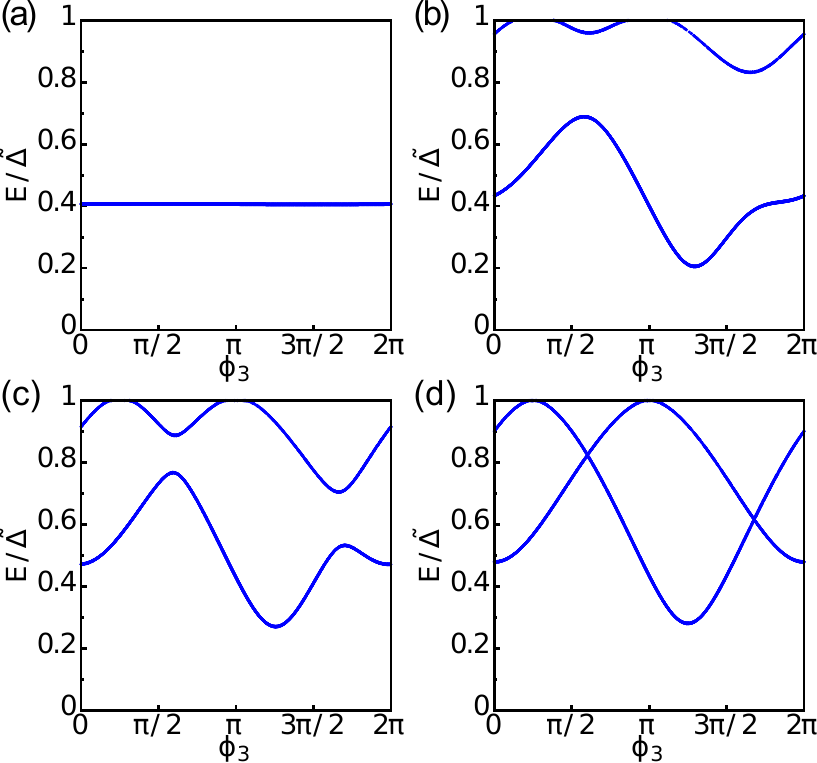}
\caption{The overview of the ABS spectrum. The ABS energies are plotted versus the phase of the third lead $\varphi_3$ for different separations $L$. For all plots, $t_{A}=0.85$, $t_{B}=0.95$, $\theta_1^A=\theta_3^B=0$, and $\theta_3^A=\theta_2^B=-\pi$, $\tau \Delta =0.2$,  $\varphi_1=\pi$, $\varphi_2=\pi/4$. (a) $L/(v_{\rm w}\tau)=0.1$. The strong 1D hybridization regime: a single ABS in both junctions hardly depending on $\varphi_3$. (b) $L/(v_{\rm w}\tau)=1$. The crossover between the regimes. The second ABS emerges from the continuous spectrum. It remains close to the band edge. (c) $L/(v_{\rm w}\tau)=2$. The system tends towards the formation of two independent ABS. The energy splitting at anticrossings is still comparable with $\tilde{\Delta}$.   (d) $L/(v_{\rm w}\tau)=6$. The weak 1D hybridization regime. Two ABS are almost independent, the energy splitting near degeneracy points is almost invisible. }
\label{fig:Ldependence}
\end{center}
\end{figure}

The crossover between the regimes is not trivial since the number of ABS in two limits are different. We illustrate the crossover by numerical calculations presented in Fig. \ref{fig:Ldependence}. In the Figure, we plot the ABS spectrum versus the phase of the third lead, $\varphi_3$, at various separations between the junctions and for representative choice of the junction scattering matrices. In Fig. \ref{fig:Ldependence} (a) that corresponds to a small separation and strong 1D hybridization regime, we observe a single ABS with no $\varphi_3$ dependence. The second ABS emerges from continuous spectrum at larger separations (Fig. \ref{fig:Ldependence} (b)), and the energies get closer to each other upon increasing $L$  (Fig. \ref{fig:Ldependence} (c)). Deep in the weak 1D hybridization regime, the ABS energies correspond to independent junction states with virtually unvisible anticrossings (Fig. \ref{fig:Ldependence} (d)). The emergence of the second ABS from the continuum is of separate interest and is investigated in Sec. \ref{sec:Transition}.

A clear idealized case is where the propagation in the junctions is ballistic like in the covered sections of the nanowire. In principle, this can be realized in sufficiently pure nanowires.
This case is characterized by the absence of quantum interference involving the phase $k_{\rm F}L$ since the electrons or holes are never reflected, and zero-energy crossings of ABS. It is detailed in Sec. \ref{sec:Transparent}.

In general, the junctions are not transparent, that is, $t_{A,B}\neq 1$, the  electrons and holes propagaring between the junctions may reflect from those and bounce in the piece of the nanowire covered by the third lead. The bounces result in the quantum interference pattern involving the phase $k_{\rm F}L$. This pattern can be observed experimentally by changing $k_{\rm F}$ slightly with a back gate. We discuss and illustrate the interference in Sec. \ref{sec:Interference}. It is clearly visible in both 1D regimes.

If the 1D propagation ampitudes become sufficiently small, $\simeq {G_{\rm Q} R}$, we enter the competition regime (Fig. \ref{fig:regimes}). 
To describe this, we extend the perturbation theory of Sec. \ref{sec:LongLead} to include the 3D propagation amplitudes next to the 1D propagation amplitudes. This analysis is rather involved since 3D propagation also encompasses the electron-hole and hole-electron conversion, and is detailed in Sec. \ref{sec:Interplay}. We will show that the result can be regarded as interference of 2 independent 3D amplitudes affected by mesoscopic fluctuations in the lead and a single 1D amplitude affected by the phase $k_{\rm F}L$. To describe the 3D amplitudes, we refine the semiclassical approach suggested in Ref. \onlinecite{kornich:prr19} and eventually correct an error in that reference.

\section{Strong 1D hybridization}
\label{sec:ShortLead}
In this Section, we consider the limit $L\ll v_{\rm w}\tau, \xi_{\rm w}$, when electrons do not exhibit Andreev reflection in the piece of the nanowire covered by the third lead. For the scattering amplitudes defined in Eqs. (\ref{eq:scattering_amplitudes})-(\ref{eq:transmissiont}) this implies $r_3 \to 0$, $t \to 1$. Solving the Eq. (\ref{eq:det}) in this limit, we obtain an equation for the ABS energy,
\begin{equation}
\label{eq:ShortLead}
\sin^2{\chi}=T_{\rm s}\sin^2{\left[\frac{\varphi_1-\varphi_2}{2}\right]}.
\end{equation} 
Here, $T_{\rm s}$ is in fact the transmission coefficient  of the normal scattering in a compound junction obtained by putting the junctions $A$ and $B$ in series. It is given by the usual expression (see, e.g. Ref. \onlinecite{Transport})
\begin{equation}
\label{eq:Ts}
T_{\rm s}=\frac{t_A^2t_B^2}{1 + r_A^2r_B^2+2r_Ar_B\cos{\theta}},\end{equation}
where $\theta \equiv \theta_3^A+\theta_3^B - 2k_{\rm F}L$.
As a rather trivial interference effect, it involves the phase accumulated in the course of round trip between the junctions.

As mentioned in the Section \ref{sec:WaveFunctions}, the dependence of $\chi$ on the parameter $\tau \Delta$ is insignificant if normalized on the proximity gap $\tilde{\Delta}$. So we can approximate $\sin \chi \approx \sqrt{1 -(E/\tilde{\Delta})^2}$. This reproduces a standard relation for an ABS in a one-channel junction between two leads \cite{Transport}:
\begin{equation}
E_{\rm ABS} = \tilde{\Delta} \sqrt{1 - T_{\rm s} \sin^2 {\left[\frac{\varphi_1-\varphi_2}{2}\right]}}.
\end{equation}

\section{Weak 1D hybridization: perturbation theory}
\label{sec:LongLead}

Let us turn to the opposite limit $L/\xi_{\rm w} \gg 1$. In this weak 1D hybridization regime, the transmission amplitude $t$ is small, eventually, exponentially small, $t = (1-e^{2i\chi})e^{-L/\xi_{\rm w}}$. We will develop a perturbation theory for the ABS energies in terms of $t$. We restrict ourselves to the most important situation of the vicinity of the degeneracy points, where the energies of two ABS formed at the junctions $A$ and $B$, almost coincide. The perturbation lifts the degeneracy resulting in the anticrossing of two energy levels. The energy splitting at the anticrossing $\delta E$ is much smaller than $\tilde{\Delta}$, $\delta E \simeq |t| \tilde{\Delta}$. 

The derivation is as follows. 
In the limit $t=0$ the scattering matrix $S_{\rm NS}S_{\rm AS}$ is separated into two independent $4\times4$ blocks corresponding to the junctions $A$ and $B$. We examine the eigenvectors of the blocks and pick up one corresponding to the eigenvalue $1$ at certain energy, that is, to the ABS energy. 
The perturbation enters an off-diagonal $4\times4$ block. We project this block on the eigenvectors $|A\rangle$ and $|B\rangle$ found for the $A$ and $B$ blocks. 
We take the derivative of the diagonal blocks $A$ and $B$ with respect to energy. With this, we obtain an  effective $2\times 2$ Hamiltonian to describe the anticrossing region,
\begin{eqnarray}
\label{eq:perturbationHamiltonian}
H_{\rm eff}=E_0 + \begin{pmatrix}
\delta E_A & {\cal M}\\
{\cal M}^* & \delta E_B
\end{pmatrix},
\end{eqnarray}
where $E_0$ is the energy at the degeneracy point, $\delta E_{A,B}$ are small deviations from the degeneracy in zeroth order in $|t|$, and ${\cal M} \propto t$ is the non-diagonal matrix element representing the perturbation. This element contains the expressions for the 4-eigenvectors that are rather clumsy. In the most compact form, it can be expressed using the notations 
\begin{eqnarray}
\sqrt{2} u_{A,B}^\pm=\sqrt{1\pm{\rm sgn}\tilde{\varphi}_{1,2}\sqrt{1-\frac{r_{A,B}^2}{\cos^2{\chi_0}}}},
\end{eqnarray}
$(u^+_{A,B})^2 + (u^-_{A,B})^2 =1$, $u^{\pm}$ are related to electron and hole amplitudes in the third lead.
The matrix element is defined upon an arbitrary phase factor and reads 
\begin{eqnarray}
\label{eq:calM}
{\cal M}=\frac{e^{-L/\xi_{\rm w}}\sin{\chi_0}}{\chi'(E_0)}\left[u_B^-u_A^+e^{-i\theta/2}-u_A^-u_B^+e^{i\theta/2}\right],
\end{eqnarray}
where $\chi'(E_0)=\partial\chi/\partial E|_{E=E_0}$, $\chi_0=\chi(E_0)$.

The matrix element is thus contributed by two amplitudes corresponding to the right- and left-moving electrons. If the junctions are ballistic, only one of these amplitudes survives depending on the ${\rm sgn}\tilde{\varphi_1}$ (${\rm sgn}\tilde{\varphi}_2=-{\rm sgn}\tilde{\varphi}_1$ in the anticrossing). This case is further detailed in Sec. \ref{sec:Transparent}.

The energy splitting then assumes the form
\begin{eqnarray}
\label{eq:EnergySplitting}
\delta E^2=4 |{\cal M}|^2 = C((u_A^+u_B^-)^2+(u_A^-u_B^+)^2-\\ \nonumber-2u_A^-u_A^+u_B^-u_B^+\cos\theta)),\ \ \ 
\end{eqnarray}
where
\begin{eqnarray}
C&=&\frac{4 e^{-2 L/\xi_{\rm w}}\sin^2{\chi_0}}{(\chi'(E_0))^2}.
\end{eqnarray} 
If we implement the heuristic approximation we made for $\chi(E)$, $C = 4 (\tilde {\Delta}-E^2/\tilde{\Delta})^2 e^{-2 L/\xi_{\rm w}}$.

The Eq. (\ref{eq:EnergySplitting}) makes explicit the interference pattern that is periodic in $\theta$. Moreover, both amplitudes become equal in modulus and the energy splitting vanishes at $\theta = 0$ provided the junctions have the same transmission coefficients and ${\rm sgn}\tilde{\varphi}_1 = {\rm sgn}\tilde{\varphi}_2$. 

\section{Interference at $L \simeq \xi_{\rm w}$}
\label{sec:Interference}

\begin{figure}[tb]
\begin{center}
\includegraphics[width=\linewidth]{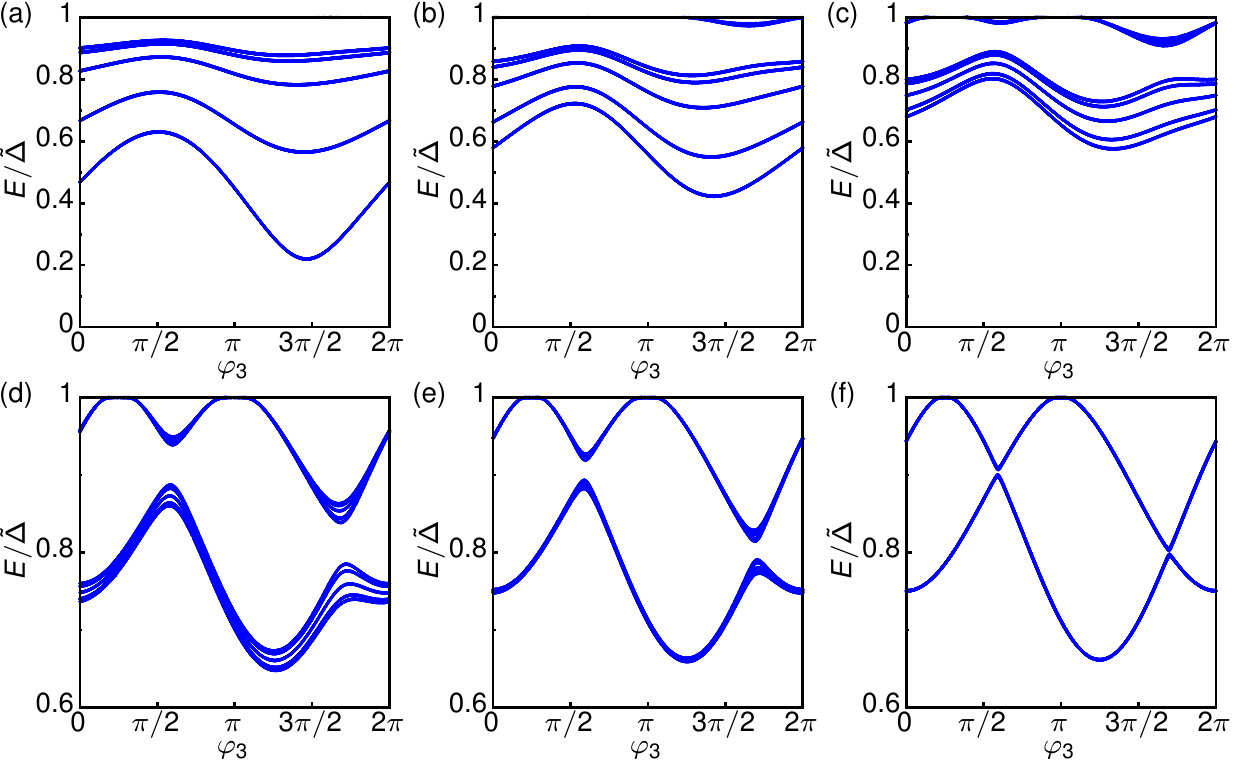}
\caption{The overview of the interference effect. The ABS energies at different settings of the phase $k_{\rm F} L$ versus the phase of the third lead $\varphi_3$. The values of the separation for the subplots are: $L/(v_{\rm w}\tau) =$ (a) $0.25$, (b) $0.50$, (c) $1.00$, (d) $2.00$, (e) $3.00$, (f) $5.00$. In each subplot, the accumulated phase takes the value $k_{\rm F} L\  {\rm mod}\ \pi= \{0,1,2,3,4\} \pi/8$, and the curves move upwards upon increasing the phase. For all the plots $t_{A}=0.6$, $t_{B}=0.7$, $\theta_1^A=\theta_3^B=0$, $\theta_3^A=\theta_2^B=-\pi$, $\tau \Delta =0.2$, $\varphi_1=\pi$, $\varphi_2=\pi/4$.
(a) The strong 1D hybridization regime, a single ABS, the interference modulates the transmission coefficient of the transparent junction. (b) The second ABS appears, the interference effect is still strong over the whole range of $\varphi_3$. (c)-(e) The effect is gradually confined to the anticrossing regions. (f) The weak 1D hybridization regime, the energy splitting near degeneracy points is not visible although is still affected by the interference. }
\label{fig:interference}
\end{center}
\end{figure}

In both regimes of strong and weak 1D hybridization, we have seen a significant interference effect, Eqs. (\ref{eq:Ts}), (\ref{eq:EnergySplitting}). However, in the strong hybridization regime the effect was confined to 
the ABS energies not depending on the phase of the third lead, while in the weak hybridization regime it was visible in the vicinity of the degeneracy points only. This motivates us to explore the effect at the intermediate values of $L \simeq \xi_{\rm w}$. The numerical results obtained are presented in Fig. \ref{fig:interference}. The subplots are computed at increasing values of $L$. In each subplot, the different curves correspond to different values of the phase $k_{\rm F} L$.

As we see, the significant interference effect is compatible with $\varphi_3$-dependence of the curves, that is, with significant probability of Andreev reflection between the junctions. However, its magnitude at arbitrary gradually reduces upon increasing $L$ and becomes confined to anticrossing regions at $L \simeq 3 v_{\rm w} \tau$. 

\begin{figure}[tb]
\begin{center}
\includegraphics[width=0.8\linewidth]{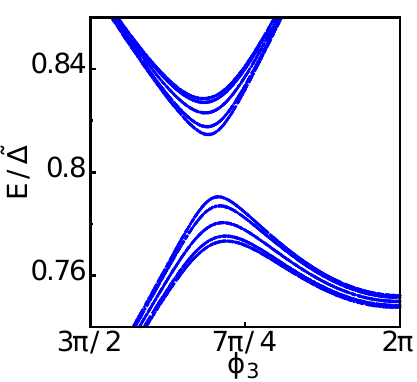}
\caption{A zoom of an anticrossing region in Fig. \ref{fig:interference} (e). The energies are computed numerically and coincide with the perturbation theory results of Sec. \ref{sec:LongLead} in three significant digits. }
\label{fig:interferencezoom}
\end{center}
\end{figure}

In Fig. \ref{fig:interferencezoom} we present the zoom on the vicinity of the degeneracy point, this makes the strong interference effect evident. For this parameter choice, the spectrum in the zoom window is described by the perturbation Hamiltonian (\ref{eq:perturbationHamiltonian}) with the accuracy of 3 significant digits.  

\section{Upper ABS merging with the continuum} 
\label{sec:Transition} 
Generally, an upper ABS that persist in a multi-terminal at certain phase settings, may disappear merging with the continuous spectrum. In a general context, this situation  has been thoroughly investigated in Ref. \onlinecite{Yokoyama}.  For our three-terminal setup with no appreciable spin-orbit interaction, this consideration predicts the gap edge touching (GET) curves in the two-dimensional space of the phases $\tilde{\varphi}_1$, $\tilde{\varphi}_2$. The merging occurs at these curves.

Our setup provides a natural cause for such merging since we expect a single ABS in the strong 1D hybridization regime and two ABS in the weak 1D hybridization regime. The upper band should therefore go to the continuum upon decreasing the separation $L$. We investigate this in detail in this Section.

It turns out that the upper ABS is present in the structure at any settings of $L$ and junction scattering matrices. However, the region in the two-dimensional space of phases where the state is present, shrinks to a line in the strong hybridization regime and fills almost the entire space in the weak hybridization regime.

\begin{figure}[tb]
\begin{center}
\includegraphics[width=\linewidth]{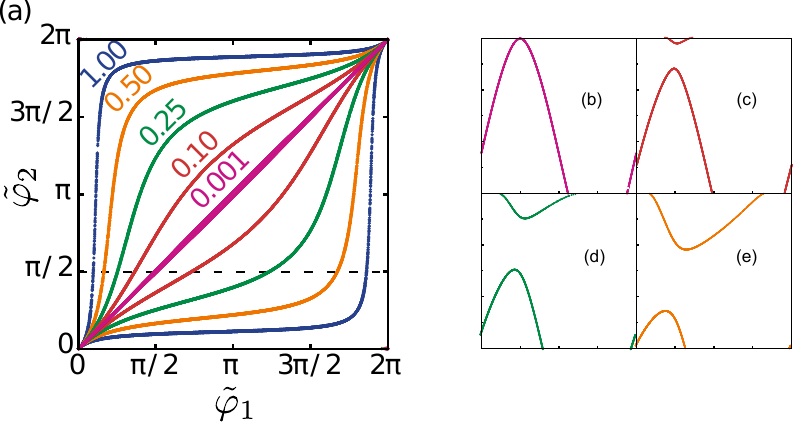}
\caption{
Gap edge touching by the upper ABS. 
(a) The GET curves in the plane $(\tilde{\varphi}_1,\tilde{\varphi_2})$ for different separations $L/(v_{\rm w}\tau)$ given in the labels. The fixed parametes are: $t_{A}=0.85$, $t_{B}=0.95$, $\tau \Delta =0.2$, $v_{F}/(L\Delta) =1$, $\varphi_3=0$, $\theta_1^A=\theta_3^B=0$, $\theta_3^A=\theta_2^B=-\pi$, and $k_{\rm F}L=\pi/4$. (b)-(e) The ABS energies at $\tilde{\varphi}_2=\pi/4$ (dashed line in (a)) illustrate the merging of the upper ABS with the continuous spectrum. 
The values of the separation go through $L/(v_{\rm w}\tau)=\{0.001, 0.1,0.25, 0.5\}$ from (b) to (e). For all plots, the vertical axis is $E/\tilde{\Delta}$ ranging from 0.7 to 1, the horizontal axis is $\tilde{\varphi}_1$ ranging from 0 to $2\pi$.}
\label{fig:PocketsEnergy}
\end{center}
\end{figure}

This is illustrated in Fig. \ref{fig:PocketsEnergy}(a) where we plot the GET curves for various $L$ in an elementary cell $(0,0), (2\pi, 2\pi)$ (The overall spectrum is periodic in both phases with the period $2\pi$). The curves are symmetric with respect to $\tilde{\varphi}_1 = \tilde{\varphi}_2$ line. At vanishing $L$, the curves converge to the line. It is easy to understand why. Since the third lead is irrelevant, there is a zero phase difference at this line for the resulting 2-terminal junction. It is known to be a GET point for a two-terminal junctions \cite{Transport}. Upon increasing $L$, the curves move apart bounding a region where the upper ABS is present. Already at $L/(v_{\rm w}\tau)=1$, this region fills the elementary cell almost entirely. Upon further increase, the GET curves are pressed to the boundaries of the elementary cell where either $\tilde{\varphi}_1 =0$ or $\tilde{\varphi}_2 =0$. Indeed, in this limit we have two independent two-terminal junctions, and this defines the positions of their GET points. 

It is interesting and instructive to look at the spectrum of both ABS. It is plotted in Figs.  \ref{fig:PocketsEnergy} (b)-(e) along the line $\tilde{\varphi}_1 = \pi/2$. The subfigures correspond to different settings of $L$. The Fig. \ref{fig:PocketsEnergy} (b) corresponding to the smallest $L$ represents the {\it lowermost} ABS and seems to touch the edge at $\tilde{\varphi}_1 = \tilde{\varphi}_2$. However, it only seems. In fact, there is a tiny region near this point where the upper ABS is present, and it is separated in energy from the lowermost one. This structure becomes apparent upon increase of $L$ (see Figs. \ref{fig:PocketsEnergy} (c)-(e)).

 \section{Ballistic junctions}
 \label{sec:Transparent}
 In this Section we concentrate on the special case of ballistic junctions, implying no normal reflection in the regions $A,B$: $r_A = r_B=0$. The spectrum separates into two parts: for right-moving electrons and left-moving holes, and for left-moving electrons and right-moving holes, that are obtained from each other by exchange of the electrons and holes. An energy level at $E$ in one part corresponds to the energy level at $-E$ in another part by virture of Bogoliubov-de Gennes symmetry. Correspondingly, the Eq. (\ref{eq:det}) splits into two parts.  The part for right-moving electrons and left-moving holes reads
\begin{eqnarray}
\label{eq:TransparentEq}
\left[e^{-i(2\chi-\tilde{\varphi}_1)}-\kappa (e^{i\tilde{\varphi}_1}-1) - 1\right]\times
\\ \nonumber
\left[e^{-i(2\chi+\tilde{\varphi}_2)}-\kappa (e^{-i\tilde{\varphi}_2}-1) - 1\right]\\ \nonumber= - 4\kappa\sin^2\chi.
\end{eqnarray}
Here, $\kappa \equiv \exp(-2L/\xi_{\rm w})$ This equation is to be solved for $\chi$ and then energy for any given $\tilde{\varphi}_{1,2}$.

To understand the qualitative characteristics of the spectrum, let us consider the weak hybridization regime $\kappa \to 0$. In zeroth order approximation, two first brackets give rise to two solutions $\chi=\tilde{\varphi}_1/2$ and $\chi=\pi-\tilde{\varphi}_2/2$. Under heuristic approximation discussed, this gives rise to two ABS energies $E = \tilde{\Delta} \cos(\tilde{\varphi}_1/2)$ and $E = -\tilde{\Delta} \cos(\tilde{\varphi}_2/2)$ for the states localized at the junctions $A$ and $B$, respectively. The energies of the states cross zero at $\tilde{\varphi}_{1,2} = \pi$, which is a known peculiarity of the completely ballistic two-terminal junction \cite{Transport}. The small $\kappa$ is relevant 
at the degeneracy line $\tilde{\varphi}_1+\tilde{\varphi}_2=2\pi$ and especially near the point $\tilde{\varphi}_{1} =\tilde{\varphi}_2 = \pi$ where the degeneracy occurs at zero energy. We expand all the phases in the vicinity of this point, $\chi = \pi/2 + E \chi'(0)$, $\tilde{\varphi}_{1,2} = \pi + \delta \varphi_{1,2}$. With this, the equation reduces to 
\begin{eqnarray}
(2E \chi'(0)-\delta\varphi_1)(2E \chi'(0)+\delta\varphi_2)=4\kappa.
\end{eqnarray}
In the limit $L\rightarrow \infty$ this equation decouples into two brackets, each corresponding to junctions $A$ and $B$. Assuming, $L$ is large, but finite, we obtain
\begin{eqnarray}
\label{eq:transparentanticrossing}
E\chi'(0) =\frac{1}{4}\left[\delta\varphi_1-\delta\varphi_2\pm\sqrt{(\delta\varphi_1+\delta\varphi_2)^2+16\kappa} \right].\ \ \ \ \ \ 
\end{eqnarray}

We see that the finite hybridization removes the degeneracy at $\delta\varphi_2 = - \delta \varphi_1$. However, it does not remove the zero energy crossings. Those are just shifted to a hyperbola $\delta \varphi_1 \delta \varphi_2 + 4 \kappa=0$.

\begin{figure}[tb]
\begin{center}
\includegraphics[width=\linewidth]{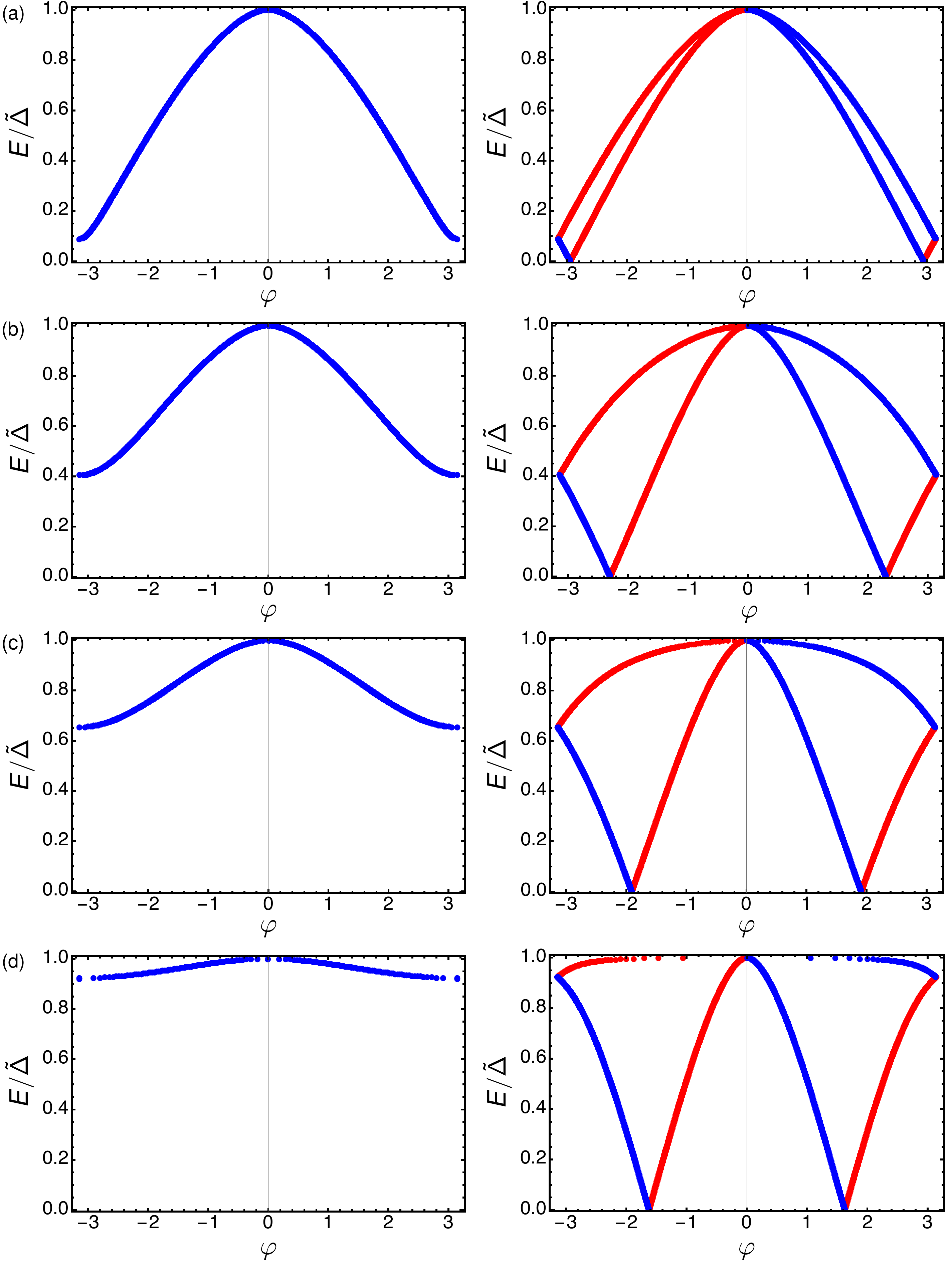}
\caption{The ABS energies for purely ballistic junctions.  We plot along the lines $\tilde{\varphi}_1=\tilde{\varphi}_2=\varphi$ (left column) and $\tilde{\varphi}_2=-\tilde{\varphi}_1=\varphi$ (right column). The energies are doubly degenerate in the left column plots. In the right column, the blue (red) color corresponds to right-(left-)moving electrons. The values of $L/(v_{\rm w}\tau)$ for the rows are: (a) $2.30$, (b) $0.8$, (c) $0.35$ (d) $0.05$. We have taken the limit $\tau\Delta \to 0$ disregarding the energy dependence of $\xi_{\rm w}$. The zero energy crossings visible in the right column is the main peculiarity of the purely ballistic case.}
\label{fig:TransparentSpectrum}
\end{center}
\end{figure}

To get an overview of the spectrum for the whole range of $L$, we plot the energies of ABS along the symmetry line $\tilde{\varphi}_1=\tilde{\varphi}_2$ (Fig. \ref{fig:TransparentSpectrum}, left column) and in the perpendicular direction
 $\tilde{\varphi}_1=-\tilde{\varphi}_2$. Along both lines, there is a convenient opportunity to make implicit plots expressing the phases through the energy. 
 
At the symmetry line, the ABS is double-degenerate: the states for right- and left-moving electrons have the same energy.
In the weak 1D hybridization regime (Fig. \ref{fig:TransparentSpectrum}(a)), the phase dependence approaches that of independent junctions. However, in accordance with Eq. (\ref{eq:transparentanticrossing}), the zero-energy crossing is shifted from the symmetry line even for small $\kappa$. Upon decreasing $L$, (Figs. \ref{fig:TransparentSpectrum}(b)-(d)), the energy raises approaching the gap egde, this is in accordance with the limit of a single compound junction. 

For the plots in the perpendicular direction, the curves of blue (red) color correspond to right- (left-)moving electrons. We see the energy crossings that is a hallmark of the ballistic junction case. The positions of the crossing gradually shift from $\pm\pi$ at big separations to $\pm \pi/2$ at small separations in accordance with the limits of independent junctions and a single compound junction.

\begin{figure}[tb]
\begin{center}
\includegraphics[width=\linewidth]{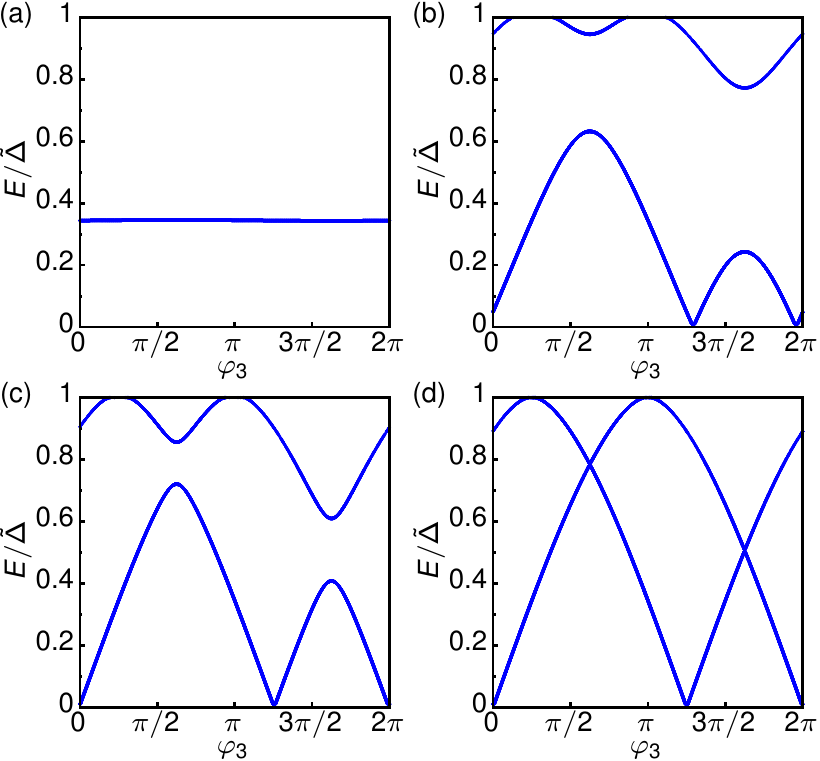}
\caption{The ABS for the setup with purely ballistic junctions ($r_A=r_B=0$) versus the phase of the third lead $\varphi_3$ for a set of different separations $L$. All parameters except $t_{A,B}$ are the same as for the plots in Fig. \ref{fig:Ldependence}. }
\label{fig:Ldependence_transparent}
\end{center}
\end{figure}

We remind that there is no interference effect on ABS since there is no normal scattering at the junctions. The plots along the lines $\tilde{\varphi}_1= \pm \tilde{\varphi}_2$ do not visually resemble those in Fig. \ref{fig:Ldependence} which may lead to the idea that the spectra are very different. To prevent this, we replot the ABS for ballistic case in Fig. \ref{fig:Ldependence_transparent} for the same parameters except setting $r_A=r_B=0$. The resulting plots do resemble those in Fig. \ref{fig:Ldependence}, zero-energy crossings being the only qualitative difference. 

\section{Competition between 1D and 3D propagation} 
\label{sec:Interplay}
In this Section, we consider the competition of 1D and 3D electron propagation as seen in the hybridization of the ABS in the Andreev molecule setup under consideration. As we have seen, the 1D propagation amplitudes $t^{e,h}_{R,L}$ between the junctions formally become exponentially small. However, this should not immediatedy imply the exponentially small hybridization. As estimated in Ref. \onlinecite{kornich:prr19}, the 3D propagation amplitudes are of the order of $\sqrt{RG_{\rm Q}}$, $R$ being a resistance characterizing the lead, and thus are not exponentially small provided the separation $L \lesssim \xi_{\rm s}$. 

A full and simultaneous account for 1D and 3D propagation seems a formidable task. In principle, it can be achieved by a non-local extension of the self-energy in Eq. (\ref{eq:SelfEnergy}): $\Sigma(x) \to \Sigma(x,x')$. However, such self-energy cannot be conveniently averaged over the disorder in the superconducting lead without cancelling the effect, which makes it hardly computable. A solution could be brute-force numerical computation of the Green's function for an atomic-level lattice model. However, such numerical exercises are seldom conclusive in practise, in view of long computation times and arbitrary modelling.

We proceed with a different method which may seem heuristic, but, in fact, is completely adequate to the problem in hand.  To explain it, let us formulate a problem in terms of scattering matrix for the junctions. Whatever the propagation, it can be incorporated into (electron and hole) transmission amplitudes between the junctions. Let us note that the competition occurs for small amplitudes where a perturbation theory is applicable. In this case, the amplitudes can be regarded as the sums over possible electron trajectories connecting the junctions. There is a direct 1D trajectory that connects the junctions through the nanowire. It accounts for the amplitudes $t^{e,h}_{R,L}$ considered above. In addition, there are trajectories where an electron starts at the junction, escapes to the lead at rather short distances $v_{\rm w} \tau \ll L$, travels in the lead, and returns to the nanowire close to the opposite junction. In distinction from the 1D amplitude, the 3D amplitude represented by the sum over these trajectories is a random quantity: it depends on the disorder configuration in the superconducting lead and vanishes upon the averaging over disorder. Importantly, the variance of this amplitude can be averaged over disorder and is determined by the properties of the superconducting lead at the space scale $L$ rather than the details of the escape. Technically, it is computed as the average of electron Green's function $G({\bf r}, {\bf r'})$, ${\bf r}$, ${\bf r}'$ being close to the opposite junctions. Besides, there are trajectories that enter and escape the nanowire several times. Since the wire is separated from the lead by a tunnel barrier, and the wire cross-section is small compared to that of the lead, the contribution of such trajectories can be safely neglected. In conclusion, the relevant transmission amplitude in the competition regime is a sum of the 1D amplitude specified above, and a random 3D amplitude. Let us compute the hybridization. 

First of all, we need to extend the perturbation theory developed in Sec. \ref{sec:LongLead} onto arbitrary set of transmission amplitudes connecting the junctions $A$, $B$. The non-diagonal matrix element ${\cal M}$ can be presented in the following form (cf. Eq. (\ref{eq:calM}))
\begin{eqnarray}
\label{eq:delta}
&&{\cal M}=\frac{1}{2 \chi'(E_0)}\times\\ \nonumber &&\times[e^{-i\frac{\theta_3^A-\theta_3^B}{2}}e^{i\varphi_3}t_{he}u_{A}^+u_{B}^++e^{-i\frac{\theta_3^A+\theta_3^B}{2}}t_{hh}u_{A}^+u_{B}^-+\\ \nonumber&&+e^{i\frac{\theta_3^A+\theta_3^B}{2}}t_{ee}u_{A}^-u_{B}^++e^{i\frac{\theta_3^A-\theta_3^B}{2}}e^{-i\varphi_3}t_{eh}u_{A}^-u_{B}^-].
\end{eqnarray}
The Eq. (\ref{eq:calM}) is reproduced if we leave here only the direct 1D propagation amplitudes substituting $t^{eh}=t^{he}=0$, $t^{ee} = t_{L}^e$, $t^{hh}=t_{R}^{h}$, with  $t_{L}^e$ ,$t_{R}^h$ taken from Eq. (\ref{eq:transmission}). 

We need to add the 3D amplitudes. We choose two points in the lead  ${\bm r}_A$ and ${\bm r}_B$, that are close to the corresponding junctions. The matrix of four transmission amplitudes is related to the Green's function describing the propagation between the points as follows\cite{nazarov:prl94}:
\begin{eqnarray}
\label{eq:tG}
t_{AB}=\frac{i}{2\pi\nu}G_3({\bm r}_A,{\bm r}_B),
\end{eqnarray}
$\nu$ being the density of states in the lead per one spin direction. Owing to the assumption of the uniform order parameter, the Green's function $G_3({\bm r}_A,{\bm r}_B)$ can be related to the quantum propagator $P({\bm r}_A,{\bm r}_B,\xi)$ defined in terms of the exact electron wavefunctions $\Psi_n({\bm r})$ in the normal state,
\begin{equation}
P({\bm r}_A, {\bm r}_B,\xi) = \sum_n \Psi^*_n({\bm r}_A) \Psi_n({\bm r}_B) \delta(\xi - \xi_n),
\end{equation}
and thus expressed in terms of the electron propagation in the normal state,
\begin{eqnarray}
&&G_3({\bm r}_A,{\bm r}_B)=\\ \nonumber &&=\int d\xi P({\bm r}_A,{\bm r}_B,\xi)\frac{1}{\xi^2+\Delta^2-E^2}\begin{pmatrix}E+\xi & \Delta e^{i\varphi_3}\\ \Delta e^{-i\varphi_3} & E-\xi\end{pmatrix},
\end{eqnarray}

Using Eq. (\ref{eq:tG}), we define two 3D amplitudes $A_{\rm e}$ and $A_{\rm o}$ for the diffusive case as
\begin{eqnarray}
A_{\rm e}&=&\int \frac{d\xi}{2\pi\nu} \frac{\sqrt{\Delta^2-E^2}}{\xi^2+\Delta^2-E^2}P({\bm r}_A,{\bm r}_B,\xi),\\
A_{\rm o}&=&\int \frac{d\xi}{2\pi\nu} \frac{\xi}{\xi^2+\Delta^2-E^2}P({\bm r}_A,{\bm r}_B,\xi).
\end{eqnarray}
Those are real in the subgap region $|E|<\Delta$ provided we assume time reversibility in the normal state. For the energies above the gap, $A_{\rm e}$ becomes imaginary and these two amplitudes can be related to real and imaginary parts of an electron wave at ${\bf r}_B$, that is emitted from a source at ${\bf r}_A$. With this, the transmission amplitudes are represented as the sum of the 1D propagation amplitudes and two random 3D amplitudes $A_{{\rm e},{\rm o}}$ taken with proper coefficients,
\begin{eqnarray}
t_{ee}&=&\frac{i E}{\sqrt{\Delta^2-E^2}}A_{\rm e}+ i A_{\rm o} + e^{ik_F L}(1-e^{2i\chi})e^{-L/\xi_w},\ \ \ \ \ \ \ \ \ \ \\
t_{eh}&=&\frac{i \Delta e^{i\varphi_3}}{\sqrt{\Delta^2-E^2}}A_{\rm e},\\
t_{he}&=&\frac{i \Delta e^{-i\varphi_3}}{\sqrt{\Delta^2-E^2}}A_{\rm e},\\
t_{hh}&=&\frac{iE}{\sqrt{\Delta^2-E^2}}A_{\rm e}- i A_{\rm o} + e^{-ik_F L}(1-e^{2i\chi})e^{-L/\xi_w}.
\end{eqnarray}

To obtain the variances of the random $A_{{\rm e},{\rm o}}$ we implement the relation between the product of two quantum propagators and the semiclassical propagator
$\mathcal{P}({\bm r}_A,{\bm r}_B, t)$, that gives the probability for a particle to be at the point ${\bm r}_B$ at the time moment $t$, provided it is at ${\bm r}_A$ in the time moment 0. This relation was implemented in Ref. \onlinecite{kornich:prr19} and reads 
\begin{equation}
\label{eq:final}
\frac{\nu}{2\pi} \int dt {\cal P}({\bm r}_A, {\bm r}_B,t) e^{i(\xi-\xi')t} = P({\bm r}_B,{\bm r}_A,\xi) P({\bm r}_A,{\bm r}_B,\xi').
\end{equation}
We have to admit a calculation error made in Ref. \onlinecite{kornich:prr19}. To correct, r.h.s. of  Eq. (3) of this work must be divided by $2 \pi$. 

With this, the variances are given by 
\begin{eqnarray}
&&\langle A_{\rm e}^2\rangle=\langle A_{\rm o}^2\rangle=\frac{1}{8\pi\nu}\int  \mathcal{P}({\bm r}_A,{\bm r}_B, t)e^{-2\sqrt{\Delta^2-E^2}|t|}dt,\ \ \ \ \ \ \\
&&\langle A_{\rm e} A_{\rm o}\rangle=0.
\end{eqnarray}
Simply enough, $A_{\rm e}$ and $A_{\rm o}$ are independent variables with equal variations.

There is a remarkably simple and general expression for the variances valid in the limit $L\ll\xi_{\rm s}$, that is, for the separations much smaller than the correlation length in the superconductor. In this case, we can replace the factor $e^{-2\sqrt{\Delta^2-E^2}|t|}$ with 1. Let us regard the lead in the normal state as a distributed conducting media earthed far from the points ${\bf r}_{A,B}$. Let us inject the current $I_A$ in the point ${\bf r}_A$ and measure the voltage $V_B$ at the point ${\bf r}_B$. This defines a three-point resistance $R\equiv V_B/I_A$. Considering kinetics of the semiclassical electron motion, we can express $R$ in terms of the semiclassical propagator,
\begin{eqnarray}
R=\frac{1}{2e^2\nu}\int_0^\infty  \mathcal{P}({\bm r}_A,{\bm r}_B,t)dt.
\end{eqnarray}
The variances are expressed in terms of this resistance,
\begin{eqnarray}
\langle A_{\rm e}^2\rangle=\langle A_{\rm o}^2\rangle=\frac{G_{\rm Q} R}{2}.
\end{eqnarray}
This expression does not depend on the geometry and resistivity distribution in the lead. 

To give a simple formula that describes the competition regime, let us assume $E\ll \Delta$, ballistic junctions, and ${{\rm sgn} \tilde{\varphi}_1}=-{{\rm sgn} \tilde{\varphi}_2}$. Under these assumptions, 
\begin{equation}
{\cal M} = \tilde{\Delta}\left[ -i A_{\rm o} + 2 e^{-ik_F L} e^{-L/\xi_{\rm w}}\right], 
\end{equation}
and the energy splitting is given by
\begin{equation}
(\delta E)^2 =4 \tilde{\Delta}^2 \left[ 4 e^{-2L/\xi_{\rm w}} + A^2_{\rm o} + 4 A_{\rm o} \sin(k_FL)  e^{-L/\xi_{\rm w}}\right].
\end{equation}
Let us note the presence of interference effect that was absent for 1D consideration of ballistic junctions. It arises due to the absence of momentum conservation in the course of 3D propagation. The 1D and 3D propagation provides in average the same contribution into the energy splitting provided  $e^{-L/\xi_{\rm w}} = \sqrt{G_{\rm Q} R/8}$.

In Ref.  \onlinecite{kornich:prr19} we have addressed the 
situation $L \simeq \xi_{\rm s}$ assuming a concrete model of a quasi-2D lead of width $L$, thickness $d \ll L$, and resistance per square $R_{\Box}$, ${\bf r}_A, {\bf r}_B$ being at the corners of the lead. The classical propagator in this case reads:
\begin{equation}
\mathcal{P}({\bm r}_A,{\bm r}_B,t)=\frac{1}{dL}\sqrt{\frac{1}{\pi D|t|}}\sum_{n=-\infty}^\infty(-1)^ne^{-D\frac{\pi^2}{L^2}n^2|t|},
\end{equation} 
$D$ being the diffusion coefficient, $D=(2e^2\nu dR_{\Box})^{-1}$. 

We neglect the contribution of 1D transmission and find 
from Eq. (\ref{eq:delta}) the average energy splitting 
\begin{eqnarray}
&&(\delta E)^2=\frac{1}{2(\chi'(E_0))^2} M G_{\rm Q} R_{\rm eff} F\left(\frac{L}{\xi_L}\right),\\
&&M=\frac{1}{\Delta^2-E^2}[\Delta^2+\\ \nonumber&&+ 2E\Delta[u_A^-u_A^+\cos{\theta^A_3}+u_B^-u_B^+\cos{\theta^B_3}]+\\ \nonumber&&+2u_A^+u_A^-u_B^+u_B^-(\Delta^2\cos{(\theta_3^A-\theta_3^B)}+\\ \nonumber&&+(2E^2-\Delta^2)\cos{(\theta_3^A+\theta_3^B)})],
\end{eqnarray}
where, conform to the definitions of Ref. \onlinecite{kornich:prr19} $R_{\rm eff}=R_{\Box}\xi_L/L$, $F(z)=4z/\pi\sum_{n=0}^\infty K_0((2n+1)z)$, $F(0)=1$. This generalizes Eq. (6) of that work to the case of arbitrary scattering matrices. A calculation error in Eq. (6) is corrected by dividing its r.h.s. by $\pi$. 
 
\section{Conclusions}

\label{sec:Conclusions}
In this work, we present a detailed study of the ABS spectrum in the three-terminal Andreev molecule setup concentrating on the effects of 1D propagation in the wire and on the competition of 1D and 3D propagation. We have identified several regimes for various relations of the junction separation $L$ as compared with the correlation lengths $\xi_{\rm w}, \xi_{\rm s}$ in the nanowire and in the superconducting lead. We have presented the details of ABS spectum in these regimes and discussed the crossovers between the regimes.  In particular, we have discussed the limits of weak and strong 1D hybridization, the interference effect, the emergence of the upper ABS from the continuous spectrum, and detailed the competition of 1D and 3D transmissions seen in the hybridization of the ABS. Our results facilitate the experimental realization of the setup where the peculiarities of Andreev spectrum can be used for quantum sensing and manipulation. 
 
 \begin{acknowledgments} 
We acknowledge useful discussions with A. Geresdi, H. Pothier, and especially with \c{C}. Girit and members of his team. This project has received funding from the European
Research Council (ERC) under the European Union's
Horizon 2020 research and innovation programme (grant
agreement No. 694272).
\end{acknowledgments}

\bibliographystyle{apsrev4-1}
\bibliography{nanowire_three_leads_bibliography_ver2}

\begin{thebibliography}{33}%
\makeatletter
\providecommand \@ifxundefined [1]{%
 \@ifx{#1\undefined}
}%
\providecommand \@ifnum [1]{%
 \ifnum #1\expandafter \@firstoftwo
 \else \expandafter \@secondoftwo
 \fi
}%
\providecommand \@ifx [1]{%
 \ifx #1\expandafter \@firstoftwo
 \else \expandafter \@secondoftwo
 \fi
}%
\providecommand \natexlab [1]{#1}%
\providecommand \enquote  [1]{``#1''}%
\providecommand \bibnamefont  [1]{#1}%
\providecommand \bibfnamefont [1]{#1}%
\providecommand \citenamefont [1]{#1}%
\providecommand \href@noop [0]{\@secondoftwo}%
\providecommand \href [0]{\begingroup \@sanitize@url \@href}%
\providecommand \@href[1]{\@@startlink{#1}\@@href}%
\providecommand \@@href[1]{\endgroup#1\@@endlink}%
\providecommand \@sanitize@url [0]{\catcode `\\12\catcode `\$12\catcode
  `\&12\catcode `\#12\catcode `\^12\catcode `\_12\catcode `\%12\relax}%
\providecommand \@@startlink[1]{}%
\providecommand \@@endlink[0]{}%
\providecommand \url  [0]{\begingroup\@sanitize@url \@url }%
\providecommand \@url [1]{\endgroup\@href {#1}{\urlprefix }}%
\providecommand \urlprefix  [0]{URL }%
\providecommand \Eprint [0]{\href }%
\providecommand \doibase [0]{http://dx.doi.org/}%
\providecommand \selectlanguage [0]{\@gobble}%
\providecommand \bibinfo  [0]{\@secondoftwo}%
\providecommand \bibfield  [0]{\@secondoftwo}%
\providecommand \translation [1]{[#1]}%
\providecommand \BibitemOpen [0]{}%
\providecommand \bibitemStop [0]{}%
\providecommand \bibitemNoStop [0]{.\EOS\space}%
\providecommand \EOS [0]{\spacefactor3000\relax}%
\providecommand \BibitemShut  [1]{\csname bibitem#1\endcsname}%
\let\auto@bib@innerbib\@empty
\bibitem [{\citenamefont {Lutchyn}\ \emph {et~al.}(2010)\citenamefont
  {Lutchyn}, \citenamefont {Sau},\ and\ \citenamefont
  {Das~Sarma}}]{lutchyn:prl10}%
  \BibitemOpen
  \bibfield  {author} {\bibinfo {author} {\bibfnamefont {R.~M.}\ \bibnamefont
  {Lutchyn}}, \bibinfo {author} {\bibfnamefont {J.~D.}\ \bibnamefont {Sau}}, \
  and\ \bibinfo {author} {\bibfnamefont {S.}~\bibnamefont {Das~Sarma}},\ }\href
  {\doibase 10.1103/PhysRevLett.105.077001} {\bibfield  {journal} {\bibinfo
  {journal} {Phys. Rev. Lett.}\ }\textbf {\bibinfo {volume} {105}},\ \bibinfo
  {pages} {077001} (\bibinfo {year} {2010})}\BibitemShut {NoStop}%
\bibitem [{\citenamefont {Oreg}\ \emph {et~al.}(2010)\citenamefont {Oreg},
  \citenamefont {Refael},\ and\ \citenamefont {von Oppen}}]{oreg:prl10}%
  \BibitemOpen
  \bibfield  {author} {\bibinfo {author} {\bibfnamefont {Y.}~\bibnamefont
  {Oreg}}, \bibinfo {author} {\bibfnamefont {G.}~\bibnamefont {Refael}}, \ and\
  \bibinfo {author} {\bibfnamefont {F.}~\bibnamefont {von Oppen}},\ }\href
  {\doibase 10.1103/PhysRevLett.105.177002} {\bibfield  {journal} {\bibinfo
  {journal} {Phys. Rev. Lett.}\ }\textbf {\bibinfo {volume} {105}},\ \bibinfo
  {pages} {177002} (\bibinfo {year} {2010})}\BibitemShut {NoStop}%
\bibitem [{\citenamefont {Mourik}\ \emph {et~al.}(2012)\citenamefont {Mourik},
  \citenamefont {Zuo}, \citenamefont {Frolov}, \citenamefont {Plissard},
  \citenamefont {Bakkers},\ and\ \citenamefont
  {Kouwenhoven}}]{mourik:science12}%
  \BibitemOpen
  \bibfield  {author} {\bibinfo {author} {\bibfnamefont {V.}~\bibnamefont
  {Mourik}}, \bibinfo {author} {\bibfnamefont {K.}~\bibnamefont {Zuo}},
  \bibinfo {author} {\bibfnamefont {S.~M.}\ \bibnamefont {Frolov}}, \bibinfo
  {author} {\bibfnamefont {S.~R.}\ \bibnamefont {Plissard}}, \bibinfo {author}
  {\bibfnamefont {E.~P. A.~M.}\ \bibnamefont {Bakkers}}, \ and\ \bibinfo
  {author} {\bibfnamefont {L.~P.}\ \bibnamefont {Kouwenhoven}},\ }\href
  {\doibase 10.1126/science.1222360} {\bibfield  {journal} {\bibinfo  {journal}
  {Science}\ }\textbf {\bibinfo {volume} {336}},\ \bibinfo {pages} {1003}
  (\bibinfo {year} {2012})},\ \Eprint
  {http://arxiv.org/abs/https://science.sciencemag.org/content/336/6084/1003.full.pdf}
  {https://science.sciencemag.org/content/336/6084/1003.full.pdf} \BibitemShut
  {NoStop}%
\bibitem [{\citenamefont {Alicea}(2012)}]{alicea:rpp12}%
  \BibitemOpen
  \bibfield  {author} {\bibinfo {author} {\bibfnamefont {J.}~\bibnamefont
  {Alicea}},\ }\href {\doibase 10.1088/0034-4885/75/7/076501} {\bibfield
  {journal} {\bibinfo  {journal} {Rep. Prog. Phys.}\ }\textbf {\bibinfo
  {volume} {75}},\ \bibinfo {pages} {076501} (\bibinfo {year}
  {2012})}\BibitemShut {NoStop}%
\bibitem [{\citenamefont {Janvier}\ \emph {et~al.}(2015)\citenamefont
  {Janvier}, \citenamefont {Tosi}, \citenamefont {Bretheau}, \citenamefont
  {Girit}, \citenamefont {Stern}, \citenamefont {Bertet}, \citenamefont
  {Joyez}, \citenamefont {Vion}, \citenamefont {Esteve}, \citenamefont
  {Goffman}, \citenamefont {Pothier},\ and\ \citenamefont
  {Urbina}}]{janvier:science15}%
  \BibitemOpen
  \bibfield  {author} {\bibinfo {author} {\bibfnamefont {C.}~\bibnamefont
  {Janvier}}, \bibinfo {author} {\bibfnamefont {L.}~\bibnamefont {Tosi}},
  \bibinfo {author} {\bibfnamefont {L.}~\bibnamefont {Bretheau}}, \bibinfo
  {author} {\bibfnamefont {{\c C}.~{\"O}.}\ \bibnamefont {Girit}}, \bibinfo
  {author} {\bibfnamefont {M.}~\bibnamefont {Stern}}, \bibinfo {author}
  {\bibfnamefont {P.}~\bibnamefont {Bertet}}, \bibinfo {author} {\bibfnamefont
  {P.}~\bibnamefont {Joyez}}, \bibinfo {author} {\bibfnamefont
  {D.}~\bibnamefont {Vion}}, \bibinfo {author} {\bibfnamefont {D.}~\bibnamefont
  {Esteve}}, \bibinfo {author} {\bibfnamefont {M.~F.}\ \bibnamefont {Goffman}},
  \bibinfo {author} {\bibfnamefont {H.}~\bibnamefont {Pothier}}, \ and\
  \bibinfo {author} {\bibfnamefont {C.}~\bibnamefont {Urbina}},\ }\href
  {\doibase 10.1126/science.aab2179} {\bibfield  {journal} {\bibinfo  {journal}
  {Science}\ }\textbf {\bibinfo {volume} {349}},\ \bibinfo {pages} {1199}
  (\bibinfo {year} {2015})},\ \Eprint
  {http://arxiv.org/abs/https://science.sciencemag.org/content/349/6253/1199.full.pdf}
  {https://science.sciencemag.org/content/349/6253/1199.full.pdf} \BibitemShut
  {NoStop}%
\bibitem [{\citenamefont {Deacon}\ \emph {et~al.}(2010)\citenamefont {Deacon},
  \citenamefont {Tanaka}, \citenamefont {Oiwa}, \citenamefont {Sakano},
  \citenamefont {Yoshida}, \citenamefont {Shibata}, \citenamefont {Hirakawa},\
  and\ \citenamefont {Tarucha}}]{deacon:prl10}%
  \BibitemOpen
  \bibfield  {author} {\bibinfo {author} {\bibfnamefont {R.~S.}\ \bibnamefont
  {Deacon}}, \bibinfo {author} {\bibfnamefont {Y.}~\bibnamefont {Tanaka}},
  \bibinfo {author} {\bibfnamefont {A.}~\bibnamefont {Oiwa}}, \bibinfo {author}
  {\bibfnamefont {R.}~\bibnamefont {Sakano}}, \bibinfo {author} {\bibfnamefont
  {K.}~\bibnamefont {Yoshida}}, \bibinfo {author} {\bibfnamefont
  {K.}~\bibnamefont {Shibata}}, \bibinfo {author} {\bibfnamefont
  {K.}~\bibnamefont {Hirakawa}}, \ and\ \bibinfo {author} {\bibfnamefont
  {S.}~\bibnamefont {Tarucha}},\ }\href {\doibase
  10.1103/PhysRevLett.104.076805} {\bibfield  {journal} {\bibinfo  {journal}
  {Phys. Rev. Lett.}\ }\textbf {\bibinfo {volume} {104}},\ \bibinfo {pages}
  {076805} (\bibinfo {year} {2010})}\BibitemShut {NoStop}%
\bibitem [{\citenamefont {Tosi}\ \emph {et~al.}(2019)\citenamefont {Tosi},
  \citenamefont {Metzger}, \citenamefont {Goffman}, \citenamefont {Urbina},
  \citenamefont {Pothier}, \citenamefont {Park}, \citenamefont {Yeyati},
  \citenamefont {Nyg\aa{}rd},\ and\ \citenamefont {Krogstrup}}]{tosi:prx19}%
  \BibitemOpen
  \bibfield  {author} {\bibinfo {author} {\bibfnamefont {L.}~\bibnamefont
  {Tosi}}, \bibinfo {author} {\bibfnamefont {C.}~\bibnamefont {Metzger}},
  \bibinfo {author} {\bibfnamefont {M.~F.}\ \bibnamefont {Goffman}}, \bibinfo
  {author} {\bibfnamefont {C.}~\bibnamefont {Urbina}}, \bibinfo {author}
  {\bibfnamefont {H.}~\bibnamefont {Pothier}}, \bibinfo {author} {\bibfnamefont
  {S.}~\bibnamefont {Park}}, \bibinfo {author} {\bibfnamefont {A.~L.}\
  \bibnamefont {Yeyati}}, \bibinfo {author} {\bibfnamefont {J.}~\bibnamefont
  {Nyg\aa{}rd}}, \ and\ \bibinfo {author} {\bibfnamefont {P.}~\bibnamefont
  {Krogstrup}},\ }\href {\doibase 10.1103/PhysRevX.9.011010} {\bibfield
  {journal} {\bibinfo  {journal} {Phys. Rev. X}\ }\textbf {\bibinfo {volume}
  {9}},\ \bibinfo {pages} {011010} (\bibinfo {year} {2019})}\BibitemShut
  {NoStop}%
\bibitem [{\citenamefont {Su}\ \emph {et~al.}(2017)\citenamefont {Su},
  \citenamefont {Tacla}, \citenamefont {Hocevar}, \citenamefont {Car},
  \citenamefont {Plissard}, \citenamefont {Bakkers}, \citenamefont {Daley},
  \citenamefont {Pekker},\ and\ \citenamefont {Frolov}}]{su:natcom17}%
  \BibitemOpen
  \bibfield  {author} {\bibinfo {author} {\bibfnamefont {Z.}~\bibnamefont
  {Su}}, \bibinfo {author} {\bibfnamefont {A.~B.}\ \bibnamefont {Tacla}},
  \bibinfo {author} {\bibfnamefont {M.}~\bibnamefont {Hocevar}}, \bibinfo
  {author} {\bibfnamefont {D.}~\bibnamefont {Car}}, \bibinfo {author}
  {\bibfnamefont {S.~R.}\ \bibnamefont {Plissard}}, \bibinfo {author}
  {\bibfnamefont {E.~P.}\ \bibnamefont {Bakkers}}, \bibinfo {author}
  {\bibfnamefont {A.~J.}\ \bibnamefont {Daley}}, \bibinfo {author}
  {\bibfnamefont {D.}~\bibnamefont {Pekker}}, \ and\ \bibinfo {author}
  {\bibfnamefont {S.~M.}\ \bibnamefont {Frolov}},\ }\href@noop {} {\bibfield
  {journal} {\bibinfo  {journal} {Nature communications}\ }\textbf {\bibinfo
  {volume} {8}},\ \bibinfo {pages} {585} (\bibinfo {year} {2017})}\BibitemShut
  {NoStop}%
\bibitem [{\citenamefont {Das}\ \emph {et~al.}(2012)\citenamefont {Das},
  \citenamefont {Ronen}, \citenamefont {Most}, \citenamefont {Oreg},
  \citenamefont {Heiblum},\ and\ \citenamefont {Shtrikman}}]{das:natphys12}%
  \BibitemOpen
  \bibfield  {author} {\bibinfo {author} {\bibfnamefont {A.}~\bibnamefont
  {Das}}, \bibinfo {author} {\bibfnamefont {Y.}~\bibnamefont {Ronen}}, \bibinfo
  {author} {\bibfnamefont {Y.}~\bibnamefont {Most}}, \bibinfo {author}
  {\bibfnamefont {Y.}~\bibnamefont {Oreg}}, \bibinfo {author} {\bibfnamefont
  {M.}~\bibnamefont {Heiblum}}, \ and\ \bibinfo {author} {\bibfnamefont
  {H.}~\bibnamefont {Shtrikman}},\ }\href@noop {} {\bibfield  {journal}
  {\bibinfo  {journal} {Nature Physics}\ }\textbf {\bibinfo {volume} {8}},\
  \bibinfo {pages} {887} (\bibinfo {year} {2012})}\BibitemShut {NoStop}%
\bibitem [{\citenamefont {Deng}\ \emph {et~al.}(2012)\citenamefont {Deng},
  \citenamefont {Yu}, \citenamefont {Huang}, \citenamefont {Larsson},
  \citenamefont {Caroff},\ and\ \citenamefont {Xu}}]{deng:nanolett12}%
  \BibitemOpen
  \bibfield  {author} {\bibinfo {author} {\bibfnamefont {M.}~\bibnamefont
  {Deng}}, \bibinfo {author} {\bibfnamefont {C.}~\bibnamefont {Yu}}, \bibinfo
  {author} {\bibfnamefont {G.}~\bibnamefont {Huang}}, \bibinfo {author}
  {\bibfnamefont {M.}~\bibnamefont {Larsson}}, \bibinfo {author} {\bibfnamefont
  {P.}~\bibnamefont {Caroff}}, \ and\ \bibinfo {author} {\bibfnamefont
  {H.}~\bibnamefont {Xu}},\ }\href@noop {} {\bibfield  {journal} {\bibinfo
  {journal} {Nano letters}\ }\textbf {\bibinfo {volume} {12}},\ \bibinfo
  {pages} {6414} (\bibinfo {year} {2012})}\BibitemShut {NoStop}%
\bibitem [{\citenamefont {Plissard}\ \emph {et~al.}(2013)\citenamefont
  {Plissard}, \citenamefont {van Weperen}, \citenamefont {Car}, \citenamefont
  {Verheijen}, \citenamefont {Immink}, \citenamefont {Kammhuber}, \citenamefont
  {Cornelissen}, \citenamefont {Szombati}, \citenamefont {Geresdi},
  \citenamefont {Frolov}, \citenamefont {Kouwenhoven},\ and\ \citenamefont
  {Bakkers}}]{plissard:nnano13}%
  \BibitemOpen
  \bibfield  {author} {\bibinfo {author} {\bibfnamefont {S.~R.}\ \bibnamefont
  {Plissard}}, \bibinfo {author} {\bibfnamefont {I.}~\bibnamefont {van
  Weperen}}, \bibinfo {author} {\bibfnamefont {D.}~\bibnamefont {Car}},
  \bibinfo {author} {\bibfnamefont {M.~A.}\ \bibnamefont {Verheijen}}, \bibinfo
  {author} {\bibfnamefont {G.~W.~G.}\ \bibnamefont {Immink}}, \bibinfo {author}
  {\bibfnamefont {J.}~\bibnamefont {Kammhuber}}, \bibinfo {author}
  {\bibfnamefont {L.~J.}\ \bibnamefont {Cornelissen}}, \bibinfo {author}
  {\bibfnamefont {D.~B.}\ \bibnamefont {Szombati}}, \bibinfo {author}
  {\bibfnamefont {A.}~\bibnamefont {Geresdi}}, \bibinfo {author} {\bibfnamefont
  {S.~M.}\ \bibnamefont {Frolov}}, \bibinfo {author} {\bibfnamefont {L.~P.}\
  \bibnamefont {Kouwenhoven}}, \ and\ \bibinfo {author} {\bibfnamefont {E.~P.
  A.~M.}\ \bibnamefont {Bakkers}},\ }\href
  {https://doi.org/10.1038/nnano.2013.198} {\bibfield  {journal} {\bibinfo
  {journal} {Nature Nanotechnology}\ }\textbf {\bibinfo {volume} {8}},\
  \bibinfo {pages} {859 EP } (\bibinfo {year} {2013})}\BibitemShut {NoStop}%
\bibitem [{\citenamefont {Lee}\ \emph {et~al.}(2013)\citenamefont {Lee},
  \citenamefont {Jiang}, \citenamefont {Houzet}, \citenamefont {Aguado},
  \citenamefont {Lieber},\ and\ \citenamefont {De~Franceschi}}]{lee:nnano13}%
  \BibitemOpen
  \bibfield  {author} {\bibinfo {author} {\bibfnamefont {E.~J.~H.}\
  \bibnamefont {Lee}}, \bibinfo {author} {\bibfnamefont {X.}~\bibnamefont
  {Jiang}}, \bibinfo {author} {\bibfnamefont {M.}~\bibnamefont {Houzet}},
  \bibinfo {author} {\bibfnamefont {R.}~\bibnamefont {Aguado}}, \bibinfo
  {author} {\bibfnamefont {C.~M.}\ \bibnamefont {Lieber}}, \ and\ \bibinfo
  {author} {\bibfnamefont {S.}~\bibnamefont {De~Franceschi}},\ }\href
  {https://doi.org/10.1038/nnano.2013.267} {\bibfield  {journal} {\bibinfo
  {journal} {Nature Nanotechnology}\ }\textbf {\bibinfo {volume} {9}},\
  \bibinfo {pages} {79 EP } (\bibinfo {year} {2013})}\BibitemShut {NoStop}%
\bibitem [{\citenamefont {Chang}\ \emph {et~al.}(2013)\citenamefont {Chang},
  \citenamefont {Manucharyan}, \citenamefont {Jespersen}, \citenamefont
  {Nyg\aa{}rd},\ and\ \citenamefont {Marcus}}]{chang:prl13}%
  \BibitemOpen
  \bibfield  {author} {\bibinfo {author} {\bibfnamefont {W.}~\bibnamefont
  {Chang}}, \bibinfo {author} {\bibfnamefont {V.~E.}\ \bibnamefont
  {Manucharyan}}, \bibinfo {author} {\bibfnamefont {T.~S.}\ \bibnamefont
  {Jespersen}}, \bibinfo {author} {\bibfnamefont {J.}~\bibnamefont
  {Nyg\aa{}rd}}, \ and\ \bibinfo {author} {\bibfnamefont {C.~M.}\ \bibnamefont
  {Marcus}},\ }\href {\doibase 10.1103/PhysRevLett.110.217005} {\bibfield
  {journal} {\bibinfo  {journal} {Phys. Rev. Lett.}\ }\textbf {\bibinfo
  {volume} {110}},\ \bibinfo {pages} {217005} (\bibinfo {year}
  {2013})}\BibitemShut {NoStop}%
\bibitem [{\citenamefont {Sherman}\ \emph {et~al.}(2016)\citenamefont
  {Sherman}, \citenamefont {Yodh}, \citenamefont {Albrecht}, \citenamefont
  {Nyg{\aa}rd}, \citenamefont {Krogstrup},\ and\ \citenamefont
  {Marcus}}]{sherman:nnano16}%
  \BibitemOpen
  \bibfield  {author} {\bibinfo {author} {\bibfnamefont {D.}~\bibnamefont
  {Sherman}}, \bibinfo {author} {\bibfnamefont {J.~S.}\ \bibnamefont {Yodh}},
  \bibinfo {author} {\bibfnamefont {S.~M.}\ \bibnamefont {Albrecht}}, \bibinfo
  {author} {\bibfnamefont {J.}~\bibnamefont {Nyg{\aa}rd}}, \bibinfo {author}
  {\bibfnamefont {P.}~\bibnamefont {Krogstrup}}, \ and\ \bibinfo {author}
  {\bibfnamefont {C.~M.}\ \bibnamefont {Marcus}},\ }\href
  {https://doi.org/10.1038/nnano.2016.227} {\bibfield  {journal} {\bibinfo
  {journal} {Nature Nanotechnology}\ }\textbf {\bibinfo {volume} {12}},\
  \bibinfo {pages} {212 EP } (\bibinfo {year} {2016})}\BibitemShut {NoStop}%
\bibitem [{\citenamefont {van Woerkom}\ \emph {et~al.}(2017)\citenamefont {van
  Woerkom}, \citenamefont {Proutski}, \citenamefont {van Heck}, \citenamefont
  {Bouman}, \citenamefont {V{\"a}yrynen}, \citenamefont {Glazman},
  \citenamefont {Krogstrup}, \citenamefont {Nyg{\aa}rd}, \citenamefont
  {Kouwenhoven},\ and\ \citenamefont {Geresdi}}]{vanwoerkom:natphys17}%
  \BibitemOpen
  \bibfield  {author} {\bibinfo {author} {\bibfnamefont {D.~J.}\ \bibnamefont
  {van Woerkom}}, \bibinfo {author} {\bibfnamefont {A.}~\bibnamefont
  {Proutski}}, \bibinfo {author} {\bibfnamefont {B.}~\bibnamefont {van Heck}},
  \bibinfo {author} {\bibfnamefont {D.}~\bibnamefont {Bouman}}, \bibinfo
  {author} {\bibfnamefont {J.~I.}\ \bibnamefont {V{\"a}yrynen}}, \bibinfo
  {author} {\bibfnamefont {L.~I.}\ \bibnamefont {Glazman}}, \bibinfo {author}
  {\bibfnamefont {P.}~\bibnamefont {Krogstrup}}, \bibinfo {author}
  {\bibfnamefont {J.}~\bibnamefont {Nyg{\aa}rd}}, \bibinfo {author}
  {\bibfnamefont {L.~P.}\ \bibnamefont {Kouwenhoven}}, \ and\ \bibinfo {author}
  {\bibfnamefont {A.}~\bibnamefont {Geresdi}},\ }\href
  {https://doi.org/10.1038/nphys4150} {\bibfield  {journal} {\bibinfo
  {journal} {Nature Physics}\ }\textbf {\bibinfo {volume} {13}},\ \bibinfo
  {pages} {876 EP } (\bibinfo {year} {2017})}\BibitemShut {NoStop}%
\bibitem [{\citenamefont {Goffman}\ \emph {et~al.}(2017)\citenamefont
  {Goffman}, \citenamefont {Urbina}, \citenamefont {Pothier}, \citenamefont
  {Nyg{\aa}rd}, \citenamefont {Marcus},\ and\ \citenamefont
  {Krogstrup}}]{goffman:njp17}%
  \BibitemOpen
  \bibfield  {author} {\bibinfo {author} {\bibfnamefont {M.}~\bibnamefont
  {Goffman}}, \bibinfo {author} {\bibfnamefont {C.}~\bibnamefont {Urbina}},
  \bibinfo {author} {\bibfnamefont {H.}~\bibnamefont {Pothier}}, \bibinfo
  {author} {\bibfnamefont {J.}~\bibnamefont {Nyg{\aa}rd}}, \bibinfo {author}
  {\bibfnamefont {C.~M.}\ \bibnamefont {Marcus}}, \ and\ \bibinfo {author}
  {\bibfnamefont {P.}~\bibnamefont {Krogstrup}},\ }\href@noop {} {\bibfield
  {journal} {\bibinfo  {journal} {New Journal of Physics}\ }\textbf {\bibinfo
  {volume} {19}},\ \bibinfo {pages} {092002} (\bibinfo {year}
  {2017})}\BibitemShut {NoStop}%
\bibitem [{\citenamefont {de~Lange}\ \emph {et~al.}(2015)\citenamefont
  {de~Lange}, \citenamefont {van Heck}, \citenamefont {Bruno}, \citenamefont
  {van Woerkom}, \citenamefont {Geresdi}, \citenamefont {Plissard},
  \citenamefont {Bakkers}, \citenamefont {Akhmerov},\ and\ \citenamefont
  {DiCarlo}}]{delange:prl15}%
  \BibitemOpen
  \bibfield  {author} {\bibinfo {author} {\bibfnamefont {G.}~\bibnamefont
  {de~Lange}}, \bibinfo {author} {\bibfnamefont {B.}~\bibnamefont {van Heck}},
  \bibinfo {author} {\bibfnamefont {A.}~\bibnamefont {Bruno}}, \bibinfo
  {author} {\bibfnamefont {D.~J.}\ \bibnamefont {van Woerkom}}, \bibinfo
  {author} {\bibfnamefont {A.}~\bibnamefont {Geresdi}}, \bibinfo {author}
  {\bibfnamefont {S.~R.}\ \bibnamefont {Plissard}}, \bibinfo {author}
  {\bibfnamefont {E.~P. A.~M.}\ \bibnamefont {Bakkers}}, \bibinfo {author}
  {\bibfnamefont {A.~R.}\ \bibnamefont {Akhmerov}}, \ and\ \bibinfo {author}
  {\bibfnamefont {L.}~\bibnamefont {DiCarlo}},\ }\href {\doibase
  10.1103/PhysRevLett.115.127002} {\bibfield  {journal} {\bibinfo  {journal}
  {Phys. Rev. Lett.}\ }\textbf {\bibinfo {volume} {115}},\ \bibinfo {pages}
  {127002} (\bibinfo {year} {2015})}\BibitemShut {NoStop}%
\bibitem [{\citenamefont {Larsen}\ \emph {et~al.}(2015)\citenamefont {Larsen},
  \citenamefont {Petersson}, \citenamefont {Kuemmeth}, \citenamefont
  {Jespersen}, \citenamefont {Krogstrup}, \citenamefont {Nyg\aa{}rd},\ and\
  \citenamefont {Marcus}}]{larsen:prl15}%
  \BibitemOpen
  \bibfield  {author} {\bibinfo {author} {\bibfnamefont {T.~W.}\ \bibnamefont
  {Larsen}}, \bibinfo {author} {\bibfnamefont {K.~D.}\ \bibnamefont
  {Petersson}}, \bibinfo {author} {\bibfnamefont {F.}~\bibnamefont {Kuemmeth}},
  \bibinfo {author} {\bibfnamefont {T.~S.}\ \bibnamefont {Jespersen}}, \bibinfo
  {author} {\bibfnamefont {P.}~\bibnamefont {Krogstrup}}, \bibinfo {author}
  {\bibfnamefont {J.}~\bibnamefont {Nyg\aa{}rd}}, \ and\ \bibinfo {author}
  {\bibfnamefont {C.~M.}\ \bibnamefont {Marcus}},\ }\href {\doibase
  10.1103/PhysRevLett.115.127001} {\bibfield  {journal} {\bibinfo  {journal}
  {Phys. Rev. Lett.}\ }\textbf {\bibinfo {volume} {115}},\ \bibinfo {pages}
  {127001} (\bibinfo {year} {2015})}\BibitemShut {NoStop}%
\bibitem [{\citenamefont {Hays}\ \emph {et~al.}(2018)\citenamefont {Hays},
  \citenamefont {de~Lange}, \citenamefont {Serniak}, \citenamefont {van
  Woerkom}, \citenamefont {Bouman}, \citenamefont {Krogstrup}, \citenamefont
  {Nyg\aa{}rd}, \citenamefont {Geresdi},\ and\ \citenamefont
  {Devoret}}]{hays:prl18}%
  \BibitemOpen
  \bibfield  {author} {\bibinfo {author} {\bibfnamefont {M.}~\bibnamefont
  {Hays}}, \bibinfo {author} {\bibfnamefont {G.}~\bibnamefont {de~Lange}},
  \bibinfo {author} {\bibfnamefont {K.}~\bibnamefont {Serniak}}, \bibinfo
  {author} {\bibfnamefont {D.~J.}\ \bibnamefont {van Woerkom}}, \bibinfo
  {author} {\bibfnamefont {D.}~\bibnamefont {Bouman}}, \bibinfo {author}
  {\bibfnamefont {P.}~\bibnamefont {Krogstrup}}, \bibinfo {author}
  {\bibfnamefont {J.}~\bibnamefont {Nyg\aa{}rd}}, \bibinfo {author}
  {\bibfnamefont {A.}~\bibnamefont {Geresdi}}, \ and\ \bibinfo {author}
  {\bibfnamefont {M.~H.}\ \bibnamefont {Devoret}},\ }\href {\doibase
  10.1103/PhysRevLett.121.047001} {\bibfield  {journal} {\bibinfo  {journal}
  {Phys. Rev. Lett.}\ }\textbf {\bibinfo {volume} {121}},\ \bibinfo {pages}
  {047001} (\bibinfo {year} {2018})}\BibitemShut {NoStop}%
\bibitem [{\citenamefont {Pillet}\ \emph {et~al.}(2019)\citenamefont {Pillet},
  \citenamefont {Benzoni}, \citenamefont {Griesmar}, \citenamefont {Smirr},\
  and\ \citenamefont {Girit}}]{pillet:nanolett19}%
  \BibitemOpen
  \bibfield  {author} {\bibinfo {author} {\bibfnamefont {J.~D.}\ \bibnamefont
  {Pillet}}, \bibinfo {author} {\bibfnamefont {V.}~\bibnamefont {Benzoni}},
  \bibinfo {author} {\bibfnamefont {J.}~\bibnamefont {Griesmar}}, \bibinfo
  {author} {\bibfnamefont {J.~L.}\ \bibnamefont {Smirr}}, \ and\ \bibinfo
  {author} {\bibfnamefont {{\c C}.~{\"O}.}\ \bibnamefont {Girit}},\ }\bibfield
  {booktitle} {\emph {\bibinfo {booktitle} {Nano Letters}},\ }\href {\doibase
  10.1021/acs.nanolett.9b02686} {\bibfield  {journal} {\bibinfo  {journal}
  {Nano Letters}\ }\textbf {\bibinfo {volume} {19}},\ \bibinfo {pages} {7138}
  (\bibinfo {year} {2019})}\BibitemShut {NoStop}%
\bibitem [{\citenamefont {Scher{\"u}bl}\ \emph {et~al.}(2019)\citenamefont
  {Scher{\"u}bl}, \citenamefont {P{\'a}lyi},\ and\ \citenamefont
  {Csonka}}]{scheruebl:bjn19}%
  \BibitemOpen
  \bibfield  {author} {\bibinfo {author} {\bibfnamefont {Z.}~\bibnamefont
  {Scher{\"u}bl}}, \bibinfo {author} {\bibfnamefont {A.}~\bibnamefont
  {P{\'a}lyi}}, \ and\ \bibinfo {author} {\bibfnamefont {S.}~\bibnamefont
  {Csonka}},\ }\href@noop {} {\bibfield  {journal} {\bibinfo  {journal}
  {Beilstein J. Nanotechnol.}\ }\textbf {\bibinfo {volume} {10}},\ \bibinfo
  {pages} {363} (\bibinfo {year} {2019})}\BibitemShut {NoStop}%
\bibitem [{\citenamefont {Metalidis}\ \emph {et~al.}(2010)\citenamefont
  {Metalidis}, \citenamefont {Eschrig}, \citenamefont {Grein},\ and\
  \citenamefont {Sch\"on}}]{metalidis:prb10}%
  \BibitemOpen
  \bibfield  {author} {\bibinfo {author} {\bibfnamefont {G.}~\bibnamefont
  {Metalidis}}, \bibinfo {author} {\bibfnamefont {M.}~\bibnamefont {Eschrig}},
  \bibinfo {author} {\bibfnamefont {R.}~\bibnamefont {Grein}}, \ and\ \bibinfo
  {author} {\bibfnamefont {G.}~\bibnamefont {Sch\"on}},\ }\href {\doibase
  10.1103/PhysRevB.82.180503} {\bibfield  {journal} {\bibinfo  {journal} {Phys.
  Rev. B}\ }\textbf {\bibinfo {volume} {82}},\ \bibinfo {pages} {180503}
  (\bibinfo {year} {2010})}\BibitemShut {NoStop}%
\bibitem [{\citenamefont {Kornich}\ \emph {et~al.}(2019)\citenamefont
  {Kornich}, \citenamefont {Barakov},\ and\ \citenamefont
  {Nazarov}}]{kornich:prr19}%
  \BibitemOpen
  \bibfield  {author} {\bibinfo {author} {\bibfnamefont {V.}~\bibnamefont
  {Kornich}}, \bibinfo {author} {\bibfnamefont {H.~S.}\ \bibnamefont
  {Barakov}}, \ and\ \bibinfo {author} {\bibfnamefont {Y.~V.}\ \bibnamefont
  {Nazarov}},\ }\href {\doibase 10.1103/PhysRevResearch.1.033004} {\bibfield
  {journal} {\bibinfo  {journal} {Phys. Rev. Research}\ }\textbf {\bibinfo
  {volume} {1}},\ \bibinfo {pages} {033004} (\bibinfo {year}
  {2019})}\BibitemShut {NoStop}%
\bibitem [{\citenamefont {Chang}\ \emph {et~al.}(2015)\citenamefont {Chang},
  \citenamefont {Albrecht}, \citenamefont {Jespersen}, \citenamefont
  {Kuemmeth}, \citenamefont {Krogstrup}, \citenamefont {Nyg{\aa}rd},\ and\
  \citenamefont {Marcus}}]{chang:nnano15}%
  \BibitemOpen
  \bibfield  {author} {\bibinfo {author} {\bibfnamefont {W.}~\bibnamefont
  {Chang}}, \bibinfo {author} {\bibfnamefont {S.}~\bibnamefont {Albrecht}},
  \bibinfo {author} {\bibfnamefont {T.}~\bibnamefont {Jespersen}}, \bibinfo
  {author} {\bibfnamefont {F.}~\bibnamefont {Kuemmeth}}, \bibinfo {author}
  {\bibfnamefont {P.}~\bibnamefont {Krogstrup}}, \bibinfo {author}
  {\bibfnamefont {J.}~\bibnamefont {Nyg{\aa}rd}}, \ and\ \bibinfo {author}
  {\bibfnamefont {C.~M.}\ \bibnamefont {Marcus}},\ }\href@noop {} {\bibfield
  {journal} {\bibinfo  {journal} {Nature nanotechnology}\ }\textbf {\bibinfo
  {volume} {10}},\ \bibinfo {pages} {232} (\bibinfo {year} {2015})}\BibitemShut
  {NoStop}%
\bibitem [{\citenamefont {Kjaergaard}\ \emph {et~al.}(2016)\citenamefont
  {Kjaergaard}, \citenamefont {Nichele}, \citenamefont {Suominen},
  \citenamefont {Nowak}, \citenamefont {Wimmer}, \citenamefont {Akhmerov},
  \citenamefont {Folk}, \citenamefont {Flensberg}, \citenamefont {Shabani},
  \citenamefont {Palmstr{\o}m},\ and\ \citenamefont
  {Marcus}}]{kjaergaard:ncom16}%
  \BibitemOpen
  \bibfield  {author} {\bibinfo {author} {\bibfnamefont {M.}~\bibnamefont
  {Kjaergaard}}, \bibinfo {author} {\bibfnamefont {F.}~\bibnamefont {Nichele}},
  \bibinfo {author} {\bibfnamefont {H.~J.}\ \bibnamefont {Suominen}}, \bibinfo
  {author} {\bibfnamefont {M.~P.}\ \bibnamefont {Nowak}}, \bibinfo {author}
  {\bibfnamefont {M.}~\bibnamefont {Wimmer}}, \bibinfo {author} {\bibfnamefont
  {A.~R.}\ \bibnamefont {Akhmerov}}, \bibinfo {author} {\bibfnamefont {J.~A.}\
  \bibnamefont {Folk}}, \bibinfo {author} {\bibfnamefont {K.}~\bibnamefont
  {Flensberg}}, \bibinfo {author} {\bibfnamefont {J.}~\bibnamefont {Shabani}},
  \bibinfo {author} {\bibfnamefont {C.~J.}\ \bibnamefont {Palmstr{\o}m}}, \
  and\ \bibinfo {author} {\bibfnamefont {C.~M.}\ \bibnamefont {Marcus}},\
  }\href {\doibase 10.1038/ncomms12841} {\bibfield  {journal} {\bibinfo
  {journal} {Nature Communications}\ }\textbf {\bibinfo {volume} {7}},\
  \bibinfo {pages} {12841} (\bibinfo {year} {2016})}\BibitemShut {NoStop}%
\bibitem [{\citenamefont {G{\"u}l}\ \emph {et~al.}(2017)\citenamefont
  {G{\"u}l}, \citenamefont {Zhang}, \citenamefont {de~Vries}, \citenamefont
  {van Veen}, \citenamefont {Zuo}, \citenamefont {Mourik}, \citenamefont
  {Conesa-Boj}, \citenamefont {Nowak}, \citenamefont {van Woerkom},
  \citenamefont {Quintero-P{\'e}rez}, \citenamefont {Cassidy}, \citenamefont
  {Geresdi}, \citenamefont {Koelling}, \citenamefont {Car}, \citenamefont
  {Plissard}, \citenamefont {Bakkers},\ and\ \citenamefont
  {Kouwenhoven}}]{guel:nanolett17}%
  \BibitemOpen
  \bibfield  {author} {\bibinfo {author} {\bibfnamefont {{\"O}.}~\bibnamefont
  {G{\"u}l}}, \bibinfo {author} {\bibfnamefont {H.}~\bibnamefont {Zhang}},
  \bibinfo {author} {\bibfnamefont {F.~K.}\ \bibnamefont {de~Vries}}, \bibinfo
  {author} {\bibfnamefont {J.}~\bibnamefont {van Veen}}, \bibinfo {author}
  {\bibfnamefont {K.}~\bibnamefont {Zuo}}, \bibinfo {author} {\bibfnamefont
  {V.}~\bibnamefont {Mourik}}, \bibinfo {author} {\bibfnamefont
  {S.}~\bibnamefont {Conesa-Boj}}, \bibinfo {author} {\bibfnamefont
  {M.}~\bibnamefont {Nowak}}, \bibinfo {author} {\bibfnamefont {D.~J.}\
  \bibnamefont {van Woerkom}}, \bibinfo {author} {\bibfnamefont
  {M.}~\bibnamefont {Quintero-P{\'e}rez}}, \bibinfo {author} {\bibfnamefont
  {M.~C.}\ \bibnamefont {Cassidy}}, \bibinfo {author} {\bibfnamefont
  {A.}~\bibnamefont {Geresdi}}, \bibinfo {author} {\bibfnamefont
  {S.}~\bibnamefont {Koelling}}, \bibinfo {author} {\bibfnamefont
  {D.}~\bibnamefont {Car}}, \bibinfo {author} {\bibfnamefont {S.~R.}\
  \bibnamefont {Plissard}}, \bibinfo {author} {\bibfnamefont {E.~P. A.~M.}\
  \bibnamefont {Bakkers}}, \ and\ \bibinfo {author} {\bibfnamefont {L.~P.}\
  \bibnamefont {Kouwenhoven}},\ }\bibfield  {booktitle} {\emph {\bibinfo
  {booktitle} {Nano Letters}},\ }\href {\doibase 10.1021/acs.nanolett.7b00540}
  {\bibfield  {journal} {\bibinfo  {journal} {Nano Letters}\ }\textbf {\bibinfo
  {volume} {17}},\ \bibinfo {pages} {2690} (\bibinfo {year}
  {2017})}\BibitemShut {NoStop}%
\bibitem [{\citenamefont {Sau}\ \emph {et~al.}(2010)\citenamefont {Sau},
  \citenamefont {Tewari}, \citenamefont {Lutchyn}, \citenamefont {Stanescu},\
  and\ \citenamefont {Das~Sarma}}]{sau:prb10}%
  \BibitemOpen
  \bibfield  {author} {\bibinfo {author} {\bibfnamefont {J.~D.}\ \bibnamefont
  {Sau}}, \bibinfo {author} {\bibfnamefont {S.}~\bibnamefont {Tewari}},
  \bibinfo {author} {\bibfnamefont {R.~M.}\ \bibnamefont {Lutchyn}}, \bibinfo
  {author} {\bibfnamefont {T.~D.}\ \bibnamefont {Stanescu}}, \ and\ \bibinfo
  {author} {\bibfnamefont {S.}~\bibnamefont {Das~Sarma}},\ }\href {\doibase
  10.1103/PhysRevB.82.214509} {\bibfield  {journal} {\bibinfo  {journal} {Phys.
  Rev. B}\ }\textbf {\bibinfo {volume} {82}},\ \bibinfo {pages} {214509}
  (\bibinfo {year} {2010})}\BibitemShut {NoStop}%
\bibitem [{\citenamefont {Stanescu}\ \emph {et~al.}(2011)\citenamefont
  {Stanescu}, \citenamefont {Lutchyn},\ and\ \citenamefont
  {Das~Sarma}}]{stanescu:prb11}%
  \BibitemOpen
  \bibfield  {author} {\bibinfo {author} {\bibfnamefont {T.~D.}\ \bibnamefont
  {Stanescu}}, \bibinfo {author} {\bibfnamefont {R.~M.}\ \bibnamefont
  {Lutchyn}}, \ and\ \bibinfo {author} {\bibfnamefont {S.}~\bibnamefont
  {Das~Sarma}},\ }\href {\doibase 10.1103/PhysRevB.84.144522} {\bibfield
  {journal} {\bibinfo  {journal} {Phys. Rev. B}\ }\textbf {\bibinfo {volume}
  {84}},\ \bibinfo {pages} {144522} (\bibinfo {year} {2011})}\BibitemShut
  {NoStop}%
\bibitem [{\citenamefont {Lutchyn}()}]{Lutchyn}%
  \BibitemOpen
  \bibfield  {author} {\bibinfo {author} {\bibfnamefont {R.~M.}\ \bibnamefont
  {Lutchyn}},\ }\href@noop {} {\bibinfo  {journal} {private communications}\
  }\BibitemShut {NoStop}%
\bibitem [{\citenamefont {Pillet}\ \emph {et~al.}()\citenamefont {Pillet},
  \citenamefont {Benzoni}, \citenamefont {Griesmar}, \citenamefont {Smirr},\
  and\ \citenamefont {Girit}}]{pillet:arxive}%
  \BibitemOpen
\bibfield  {journal} {  }\bibfield  {author} {\bibinfo {author} {\bibfnamefont
  {J.~D.}\ \bibnamefont {Pillet}}, \bibinfo {author} {\bibfnamefont
  {V.}~\bibnamefont {Benzoni}}, \bibinfo {author} {\bibfnamefont
  {J.}~\bibnamefont {Griesmar}}, \bibinfo {author} {\bibfnamefont {J.~L.}\
  \bibnamefont {Smirr}}, \ and\ \bibinfo {author} {\bibfnamefont {{\c
  C}.~{\"O}.}\ \bibnamefont {Girit}},\ }\href@noop {} {\bibinfo  {journal}
  {arXiv.1809.11011v1}\ }\BibitemShut {NoStop}%
\bibitem [{\citenamefont {Nazarov}\ and\ \citenamefont
  {Blanter}(2009)}]{Transport}%
  \BibitemOpen
\bibfield  {journal} {  }\bibfield  {author} {\bibinfo {author} {\bibfnamefont
  {Y.}~\bibnamefont {Nazarov}}\ and\ \bibinfo {author} {\bibfnamefont
  {Y.}~\bibnamefont {Blanter}},\ }\href@noop {} {\emph {\bibinfo {title}
  {Quantum Transport}}}\ (\bibinfo  {publisher} {Cambridge University Press},\
  \bibinfo {year} {2009})\BibitemShut {NoStop}%
\bibitem [{\citenamefont {Yokoyama}\ and\ \citenamefont
  {Nazarov}(2015)}]{Yokoyama}%
  \BibitemOpen
  \bibfield  {author} {\bibinfo {author} {\bibfnamefont {T.}~\bibnamefont
  {Yokoyama}}\ and\ \bibinfo {author} {\bibfnamefont {Y.~V.}\ \bibnamefont
  {Nazarov}},\ }\href {\doibase 10.1103/PhysRevB.92.155437} {\bibfield
  {journal} {\bibinfo  {journal} {Phys. Rev. B}\ }\textbf {\bibinfo {volume}
  {92}},\ \bibinfo {pages} {155437} (\bibinfo {year} {2015})}\BibitemShut
  {NoStop}%
\bibitem [{\citenamefont {Nazarov}(1994)}]{nazarov:prl94}%
  \BibitemOpen
  \bibfield  {author} {\bibinfo {author} {\bibfnamefont {Y.~V.}\ \bibnamefont
  {Nazarov}},\ }\href {\doibase 10.1103/PhysRevLett.73.134} {\bibfield
  {journal} {\bibinfo  {journal} {Phys. Rev. Lett.}\ }\textbf {\bibinfo
  {volume} {73}},\ \bibinfo {pages} {134} (\bibinfo {year} {1994})}\BibitemShut
  {NoStop}%
\end{thebibliography}%

\end{document}